\documentclass[fleqn, usenatbib]{mnras}
\usepackage{style} 

\title[eBOSS Quasars]{Primordial non-Gaussianity from the Completed SDSS-IV extended Baryon Oscillation Spectroscopic Survey I:  Catalogue Preparation and Systematic Mitigation}

\author[M. Rezaie et al.]{Mehdi Rezaie$^{1}$, Ashley J. Ross$^{2}$, Hee-Jong Seo$^{1}$, Eva-Maria Mueller$^{3}$,
Will J. Percival$^{4,5,6}$,\newauthor Grant Merz$^{1}$, Reza Katebi$^{1}$, Razvan C. Bunescu$^{7}$,
Julian Bautista$^{8}$,
Joel R. Brownstein$^{9}$,\newauthor
Etienne Burtin$^{10}$, Kyle Dawson$^{9}$, H\'ector Gil-Mar\'in$^{11, 12}$,
Jiamin Hou$^{13}$, Eleanor B. Lyke$^{14}$,\newauthor
Axel de la Macorra$^{15}$,
Graziano Rossi$^{16}$, Donald P. Schneider$^{17,18}$,  Pauline Zarrouk$^{10, 19}$,\newauthor and Gong-Bo Zhao$^{20, 21}$ 
\\~\\
$^{1}$Department of Physics and Astronomy, Ohio University, Athens, OH 45701, USA\\
$^{2}$Center of Cosmology and AstroParticle Physics, The Ohio State University, Columbus, OH 43210, USA\\
$^{3}$Department of Physics, University of Oxford, Denys Wilkinson Building, Keble Road, Oxford OX1 3RH, UK\\
$^{4}$Waterloo Centre for Astrophysics, University of Waterloo, Waterloo, ON N2L 3G1, Canada\\
$^{5}$Department of Physics and Astronomy, University of Waterloo, Waterloo, ON N2L 3G1, Canada\\
$^{6}$Perimeter Institute for Theoretical Physics, 31 Caroline St. North, Waterloo, ON N2L 2Y5, Canada\\
$^{7}$Department of Computer Science, College of Computing and Informatics, University of North Carolina, Charlotte, NC 28223, USA\\
$^{8}$Institute of Cosmology \& Gravitation, Dennis Sciama Building, University of Portsmouth, Portsmouth, PO1 3FX, UK \\
$^{9}$Department of Physics and Astronomy, University of Utah, 115 S. 1400 E., Salt Lake City, UT 84112, USA\\
$^{10}$IRFU,CEA, Universit\'e Paris-Saclay, F-91191 Gif-sur-Yvette, France\\
$^{11}$Institut de Ci\`encies del Cosmos, Universitat de Barcelona, ICCUB, Mart\'i i Franqu\`es 1, E08028 Barcelona, Spain\\ 
$^{12}$Institut  d'Estudis  Espacials  de  Catalunya  (IEEC),  E08034  Barcelona,  Spain\\ 
$^{13}$Max-Planck-Institut f\"ur Extraterrestrische Physik, Postfach 1312, Giessenbachstr., 85748 Garching bei M\"unchen, Germany\\
$^{14}$Department of Physics and Astronomy, University of Wyoming, Laramie, WY 82071, USA\\
$^{15}$Instituto de Fisica, Universidad Nacional Autonoma de Mexico, Apdo. Postal 20-364, Mexico\\
$^{16}$Department of Physics and Astronomy, Sejong University, Seoul 143-747, Korea\\
$^{17}$Department of Astronomy and Astrophysics, The Pennsylvania State University, University Park, PA 16802, USA\\
$^{18}$Institute for Gravitation and the Cosmos, The Pennsylvania State University, University Park, PA 16802, USA \\ 
$^{19}$Institute for Computational Cosmology, Dept. of Physics, University of Durham, South Road, Durham DH1 3LE, UK\\
$^{20}$National Astronomy Observatories, Chinese Academy of Science, Beijing, 100101, P.R. China\\
$^{21}$School of Astronomy and Space Science, University of Chinese Academy of Sciences, Beijing 100049, P.R.China
}

\begin{document}
\label{firstpage}
\pagerange{\pageref{firstpage}--\pageref{lastpage}}
\maketitle

\begin{abstract}
 We investigate the large-scale clustering of the final spectroscopic sample of quasars from the recently completed extended Baryon Oscillation Spectroscopic Survey (eBOSS). The sample contains $343708$ objects in the redshift range $0.8<z<2.2$ and $72667$ objects with redshifts $2.2<z<3.5$, covering an effective  area of $4699~{\rm deg}^{2}$. We develop a neural network-based approach to mitigate spurious fluctuations in the density field caused by spatial variations in the quality of the imaging data used to select targets for follow-up spectroscopy. Simulations are used with the same angular and radial distributions as the real data to estimate covariance matrices, perform error analyses, and assess residual systematic uncertainties. We measure the mean density contrast and cross-correlations of the eBOSS quasars against maps of potential sources of imaging systematics to address algorithm effectiveness, finding that the neural network-based approach outperforms standard linear regression. Stellar density is one of the most important sources of spurious fluctuations, and a new template constructed using data from the Gaia spacecraft provides the best match to the observed quasar clustering. The end-product from this work is a new value-added quasar catalogue with the improved weights to correct for nonlinear imaging systematic effects, which will be made public. Our quasar catalogue is used to measure the local-type primordial non-Gaussianity in our companion paper, Mueller et al. in preparation.
\end{abstract}


\section{Introduction}
\label{sec:introduction}

The statistical properties of the initial conditions of the Universe are an open problem in modern cosmology. Single-field inflationary models predict that primordial fluctuations are almost Gaussian and any deviation toward non-Gaussianity is small \citep{guth1982fluctuations, hawking1982development, starobinsky1982dynamics, bardeen1983spontaneous, acquaviva2003gauge, maldacena2003non, zaldarriaga2004non, scoccimarro2004probing, creminelli2004single}. However, some alternative inflationary models, which include more fields, generate non-Gaussian perturbations \citep{linde1997non, Maldacena_2003, bernardeau2002non, lyth2003primordial, allen1987non, kofman1988nonflat, salopek1989designing, Chen_2007}, see, e.g., \cite{Desjacques_2010} for a review. The local-type primordial non-Gaussianity is often parameterized as \citep{matarrese2000abundance,verde2000large, komatsu2001acoustic}:
\begin{equation}
\Phi = \phi + f_{\text{NL}}(\phi^{2} - <\phi^{2}>),
\end{equation}
where $\Phi$ is the primordial gravitational field, $\phi$ is a Gaussian field, and $f_{\text{NL}}$ is called the non-linear coupling constant. For a single field inflationary model, we have $f_{\rm NL} = 5(1-n_{s})/12 \sim 0.01$, where $n_{s}$ is the scalar spectral index of primordial power spectrum. Therefore, a non-zero detection of primordial non-Gaussianity, $f_{\text{NL}} \gtrsim 1$, will rule out single-field models of inflation and refine our understanding of the early universe \citep{mukhanov1981JETPL..33..532M, starobinsky1982dynamics, hawking1982development, guth1982fluctuations, allen1987non, gangui1993three, falk1992dependence}, see, e.g., \citet{alvarez2014testing} for a review.

The current state-of-the-art constraint is $f_{\rm NL} =-0.9 \pm 5.1$ at $68$ per cent confidence level from measurements of the bispectrum of cosmic microwave background anisotropies as measured by the \textit{Planck} satellite \citep{akrami2019planck}. However, limited by cosmic variance, CMB measurements cannot reach the necessary precision to differentiate between various inflationary models \citep[e.g.,][]{baumann2009probing, abazajian2016cmb}. As an alternative route to constrain $f_{\rm NL}$ is to measure the scale-dependent effect it has on the  large-scale clustering of biased tracers at lower redshift \citep{scoccimarro2004probing, dalal2008imprints, matarrese2008effect, slosar2008constraints, taruya2008signature, grossi2008mass},
\begin{equation}\label{eq:fnlk2}
   \Delta b \propto f_{\rm NL} (b-p) \frac{1}{k^{2}T(k)},
\end{equation}
where $T(k)$ is the transfer function normalized to unity for small wavenumber $k$, $b$ is the halo bias, and $p$ is a correction factor accounting for the response of the tracer to the halo gravitational field, e.g., $1.6$ for recent mergers \citep{slosar2008constraints, reid2010JCAP...07..013R}. Due to the $k^{-2}$ dependence, the biasing effect is most apparent for fluctuations with large wavelengths \citep[see, e.g.,][for a review]{alvarez2014testing}. Therefore, constraining primordial non-Gaussianity demands galaxy redshift experiments that survey a huge cosmic volume. 

Quasars are bright quasi-stellar objects which are detectable from high redshifts thanks to the brightness of their active nuclei, and are consequently the ideal candidate for studying the distribution of matter on high redshifts, and constraining primordial non-Gaussianities with large-scale structure data \citep{giannantonio2014improved, leistedt2014constraints, karagiannis2014search, agarwal2014constraining}. Using the scale-depend bias effect on galaxy power spectrum measurements, \cite{slosar2008constraints} obtained $f_{\rm NL} = 8^{+26}_{-37}$ at $68$ per cent confidence level from the Sloan Digital Sky Survey \citep[SDSS;][]{blanton2017sloan} Data Release 5 (DR5) quasar sample \citep{adelman2007fifth}. Recently, \cite{castorina2019JCAP...09..010C} used 148,659 quasars with $0.8 < z < 2.2$ from the SDSS DR14 \citep{abolfathi2018fourteenth} to obtain $-51 < f_{\rm NL} < 21$ at $95$ per cent confidence level. 

Quasar clustering measurements at large scales are sensitive to spurious density fluctuations caused by imaging properties  \citep{ross2013MNRAS.428.1116R, Ho_2015, ross2012MNRAS.424..564R, pullen2013systematic, leistedt2013estimating, leistedt2014exploiting, kalus2019map, laurent2017clustering}, and are the primary source of systematic error that impede a robust analysis of primordial non-Gaussianity. The reasons for data having systematic errors include, but are not limited to, Galactic foregrounds (such as Galactic extinction and stellar contaminations), seeing, and survey depth variations. These properties fluctuate across the survey footprint and introduce spurious variations in the observed density field of quasars. Because many of these effects have large-scale variations, they result in an excess clustering signal at large scales. For instance, \cite{pullen2013systematic} analyzed the SDSS DR6 quasar sample \citep{richards2008efficient} to find that systematic error in the sample does not allow a robust inference on \fnl. They found the $2$\% RMS fluctuations in the quasar density, while they argued that the fluctuations under $1$\%($0.6$\%) is required to measure \fnl less than 100 (10).

Conventional techniques for improving quasar clustering measurements rely on a regression analysis of observed quasar density against a set of mappable properties that describe imaging conditions during observing \citep[e.g.,][]{ross2011ameliorating, ross2012MNRAS.424..564R,delubac2016sdss, prakash2016sdss, ross2017MNRAS.464.1168R, laurent2017clustering, bautista2018sdss}. This approach has been further enhanced to account for inter-correlations among imaging variables, validated with simulations, and applied to other photometric surveys, such as the Dark Energy Survey \citep[e.g.,][]{Elvin2018PhRvD..98d2006E,Wagoner2020arXiv200910854W}. The regression analysis fits for spurious fluctuations in target quasar density and then is employed as a selection function to assign an appropriate weight to each target to eliminate such variations. The method of mode-projection removes the modes that strongly correlate with imaging templates in covariance based estimators \citep[e.g.,][]{tegmark1997PhRvD..55.5895T, leistedt2013estimating, leistedt2014exploiting, kalus2019map}. Regression and mode-projection turn out to be mathematically equivalent \citep{kalus2016}. An alternative technique uses cross-correlations of multiple tracers, or same tracer at different redshift bins, with the assumption that each sample responds differently to imaging systematics and thus cross-correlations are not affected \citep[e.g.,][]{rhodes2013arXiv1309.5388R}. For a benchmark analysis of various cleaning methods, see, e.g., \cite{Weaverdyck2020}.

The extended Baryon Oscillation Spectroscopic Survey \citep[eBOSS;][]{dawson2016sdss} is part of the fourth phase of SDSS. The eBOSS program is the final large-galaxy redshift survey proposed within SDSS aiming to measure the expansion history and energy contents of the Universe using Large-Scale Structure (LSS)  \citep{Ahumada2020ApJS}. With a series of galaxy redshift surveys, SDSS has probed the distribution of matter traced by galaxies and quasars using the dedicated 2.5m Sloan Foundation Telescope \citep{gunn20062} at the Apache Point Observatory from 1998 to 2019. During four distinct phases, SDSS collected images and spectra of thousands to millions of astronomical objects and created the largest three-dimensional map of the cosmic web to date. The first phase of SDSS led to the detection of Baryon Acoustic Oscillations (BAO) in the large-scale clustering of galaxies \citep{eisenstein2005detection}, at the same time as the 2-degree Field Galaxy Redshift Survey \citep{cole20052df}. Calibrated by CMB measurements, the BAO scale has been widely utilized as a standard ruler leading to robust distance-redshift measurements and constraints on the nature of Dark Energy with LSS surveys. The observations of the BAO signal in LSS have improved since the early detections, passing the $5\sigma$ detection threshold and now cover a wide range in redshift using only SDSS data \citep[e.g.,][]{percival2010baryon, anderson2014clustering, alam2017clustering}. 

In the fourth phase of SDSS, eBOSS adopted the same 1000-fiber spectrograph \citep{smee2013multi} from its predecessor, BOSS \citep{dawson2012baryon}, and utilized improved pipelines for redshift estimation, background subtraction, and flux calibration \citep{bolton2012spectral, hutchinson2016redshift, jensen2016spectral, bautista2017measurement}. eBOSS introduced a new class of targets for actively star-forming galaxies with strong [OII] emission lines, known as Emission-Line Galaxies \citep[ELGs;][]{raichoor2017sdss}. ELGs extend over the redshift range of $0.6<z<1.1$ \citep{Raichoor2020MNRAS.500.3254R} and fill the redshift gap between Luminous Red Galaxies \citep[LRGs;][]{prakash2016sdss} and Quasi-Stellar Objects \citep[QSOs;][]{myers2015sdss}, referred to as quasars in this manuscript.

We present a careful assessment and treatment of imaging systematic effects in the final sample of quasars \citep{lyke2020dr16qso, ross2020lss} from the eBOSS Data Release 16 \citep{Ahumada2020ApJS}, the largest sample of quasars available to date.  We improve upon a neural network-based cleaning approach, which was originally developed, validated, and applied to eBOSS-like emission-line galaxies in \cite{rezaie2020MNRAS.495.1613R}. Compared to emission-line galaxies, quasars represent a class of sparser targets for galaxy surveys, particularly for spectroscopic surveys. This paper enhances the method to deal with the sparsity of the eBOSS DR16 quasars by modeling quasar counts per pixel with the Poisson distribution. We improve neural network training to be less prone to local minima by utilizing a cyclic learning rate. We perform a comprehensive benchmark of linear and nonlinear treatments, and investigate each method effectiveness in reducing spurious fluctuations. Residual systematic errors are quantified using mean quasar density contrasts and cross-power spectra against imaging templates. For the significance of residual spurious fluctuations, we construct covariance matrices from realistic simulations. Our primary objective is to examine, quantify, and mitigate the potential sources of observational systematic error and enhance quasar power spectrum measurements for constraining primordial non-Gaussianity. This work presents a new set of value-added catalogues with the enhanced systematic weights to account for imaging systematic effects more exquisitely compared to the standard linear regression which is used in the eBOSS pipeline. The new quasar catalogue is utilized in two accompanying papers for constraining the local-type primordial non-Gaussianity \citep{mueller2020fnl} and for exploring the impact of imaging systematic error on Baryon Acoustic Oscillations \citep{merz2020bao}.

This paper is structured as follows. Section~\ref{sec:data} describes the eBOSS quasar sample and simulated datasets used in this work. Section~\ref{sec:techniques} outlines the power spectrum estimator and our strategies for the treatment and characterization of imaging systematics. In Section~\ref{sec:results}, we present our statistical tests for quantifying residual systematic errors, assess the performance of different cleaning methods, and illustrate the impact of imaging properties on the measured power spectrum of the DR16 sample and that of the simulated catalogues. Finally, we conclude with a summary of our results and their significance for constraining primordial non-Gaussianity with quasar clustering in Section~\ref{sec:conclusion}.
\section{Data}
\label{sec:data}

This section describes the final sample of quasars from the completed eBOSS DR16 dataset \citep{ross2020lss, lyke2020dr16qso} and the simulated EZmock catalogues \citep{zhao2020qso} used in our analysis. 

\subsection{eBOSS DR16 Quasars}\label{subsec:ebossqso}

The DR16 quasar sample is the final release of large-scale structure data from SDSS-IV eBOSS and provides twice the number of quasars and sky coverage over the previous Data Release DR14. The target selection of quasars for eBOSS is presented in \cite{myers2015sdss}, and only the main details are briefly summarized here. It uses optical and infrared imaging, respectively, from SDSS and WISE \citep[Wide-Field Infrared Survey Explorer;][]{wright2010wide}. The photometric data are taken in five bands \citep[\textit{u, g, r, i, z};][]{fukugita1996sloan} and calibrated to account for the Galactic dust effect using correction factors presented in \cite{schlafly2012photometric}. The ability to obtain a quasar with redshift $z > 0.9$ is improved by 20\% using the \textsc{xdqsoz} algorithm \citep{bovy2012photometric}. The magnitude selection applies the extinction-corrected flux cuts in the g and r bands, namely $g < 22$ and $r < 22$, to choose the CORE eBOSS quasars. The targeting strategy enables the observation of the Ly-$\alpha$ high-z quasars by relaxing the maximum redshift cut and reduces stellar contaminations in the sample by incorporating a mid-infrared cut. The redshifts are estimated using the \textsc{redvsblue}\footnote{\href{https://github.com/londumas/redvsblue}{https://github.com/londumas/redvsblue}} principal component analysis algorithm described in \cite{lyke2020dr16qso}, with 95\% completeness and 2\% false positive rate. The main quasar sample spans the redshift range of $0.8 < z < 2.2$ and is used for various cosmological analyses of the BAO and RSD features \citep{hou2020qso, neveux2020qso, dumas2020qso, smith2020qso}. Additionally, the high-z quasars with $2.2<z<3.5$ are employed for measuring the BAO signal in the Ly-$\alpha$ forest \citep{chabanier2019JCAP...07..017C, blom2019A&A...629A..86B, de2019baryon}.

The preparation of the large-scale clustering catalogue of quasars is outlined in \cite{ross2020lss}. As well as associating attributes and \textit{weights} per object, the process includes generating a set of unclustered synthetic objects (often referred to as the \textit{random catalogue} or \textit{randoms}) matching the expected weighted density of quasars and accounting for the radial and angular survey geometry. Standard procedures account for the veto masking, completeness cuts, fiber collision correction by nearest neighbour upweighting, redshift failure correction through weighting, and imaging systematics by linear regression against templates. The products are the tabulated coordinates of the quasars and randoms along with appropriate columns for per-object weights, separately for the North Galactic Cap (NGC) and South Galactic Cap (SGC) regions. Each weight column is intended to address a particular systematic effect in the observed data, which is discussed briefly in the following.

The physical size of a fiber limits the ability to observe a pair of quasars within $62$ \textit{arcsec} of separation. This \textit{fiber collision} effect is not random and affects quasar clustering. Up-weighting the nearest neighbor corrects for much of the large-scale effect, although small-scales remain affected and require a more complicated procedure to achieve unbiased removal \citep{hahn2017effect, bianchi2017unbiased, mohammad2020}. In our analysis we use the simple nearest neighbour upweighting scheme as we are only interested in large scales \citep[as used, for example, by][]{neveux2020qso}. 

The completeness of redshift estimation varies among the fibers across the spectrographs, with fibers near the edge having a higher rate of redshift measurement failures. The weight $w_{\rm noz}$ is determined to mitigate this systematic error by weighting by the reciprocal of the probability that each fiber accurately measures a redshift. The impact of redshift completeness on quasar clustering is investigated in \cite{hou2020qso}. 

The remaining source of systematic uncertainty is associated with the properties of the imaging data from which targets were selected including, but not limited to, stellar contamination, Galactic extinction, and inaccurate photometric calibrations. The standard method uses a multivariate linear model to find the intercorrelation between target density and imaging templates, and provides a per-object weight $w_{\rm systot}$ to reduce this systematic effect \citep{ross2012MNRAS.424..564R, bautista2018sdss}. We explain the standard treatment and how the systematic weight $w_{\rm systot}$ is obtained in \S \ref{subsec:mitigation}. Collectively, to account for all of these observational effects, each quasar and random object must be weighted by,
\begin{equation}\label{eq:wtot}
    \centering
     w = w_{\rm systot} \times ~w_{\rm noz} \times
    ~w_{\rm FKP}\times
    ~w_{\rm cp},
\end{equation}
where $w_{\rm FKP}=[1+n(z)P_{0}]^{-1}$ is the FKP weight \citep{feldman1994ApJ...426...23F} with $P_{0}=6000~{\rm Mpc}^{3}h^{-3}$ based on the expected power\footnote{This is the default value for the BAO and RSD studies with eBOSS \cite[e.g.,][]{hou2020qso, neveux2020qso}. The impact of the FKP weight on \fnl constraints is investigated in our companion paper, \cite{mueller2020fnl}.} on scales around $0.01 - 0.3~h/{\rm Mpc}$, and $n(z)$ is the number density at redshift $z$. We require completeness $> 0.5$ (for both parameters \textsc{comb\_boss} and \textsc{sector\_ssr}\footnote{These parameters describes the survey completeness and the probability of detecting an object with a good redshift in spite of other instrumental factors.}) as well as $0.8< z < 3.5$ to avoid low-quality samples. The quasar catalogue contains $343708$ objects in the redshift range $0.8<z<2.2$ and $72667$ in the redshift interval $2.2<z<3.5$, covering an effective area of $4699$ square degrees. Hereafter, we refer to the sample with $0.8 < z < 2.2$ as \textit{main}, and the one with $2.2 < z < 3.5$ is referred to as \textit{high-z}. 

\begin{figure}
    \centering
    \includegraphics[width=0.45\textwidth]{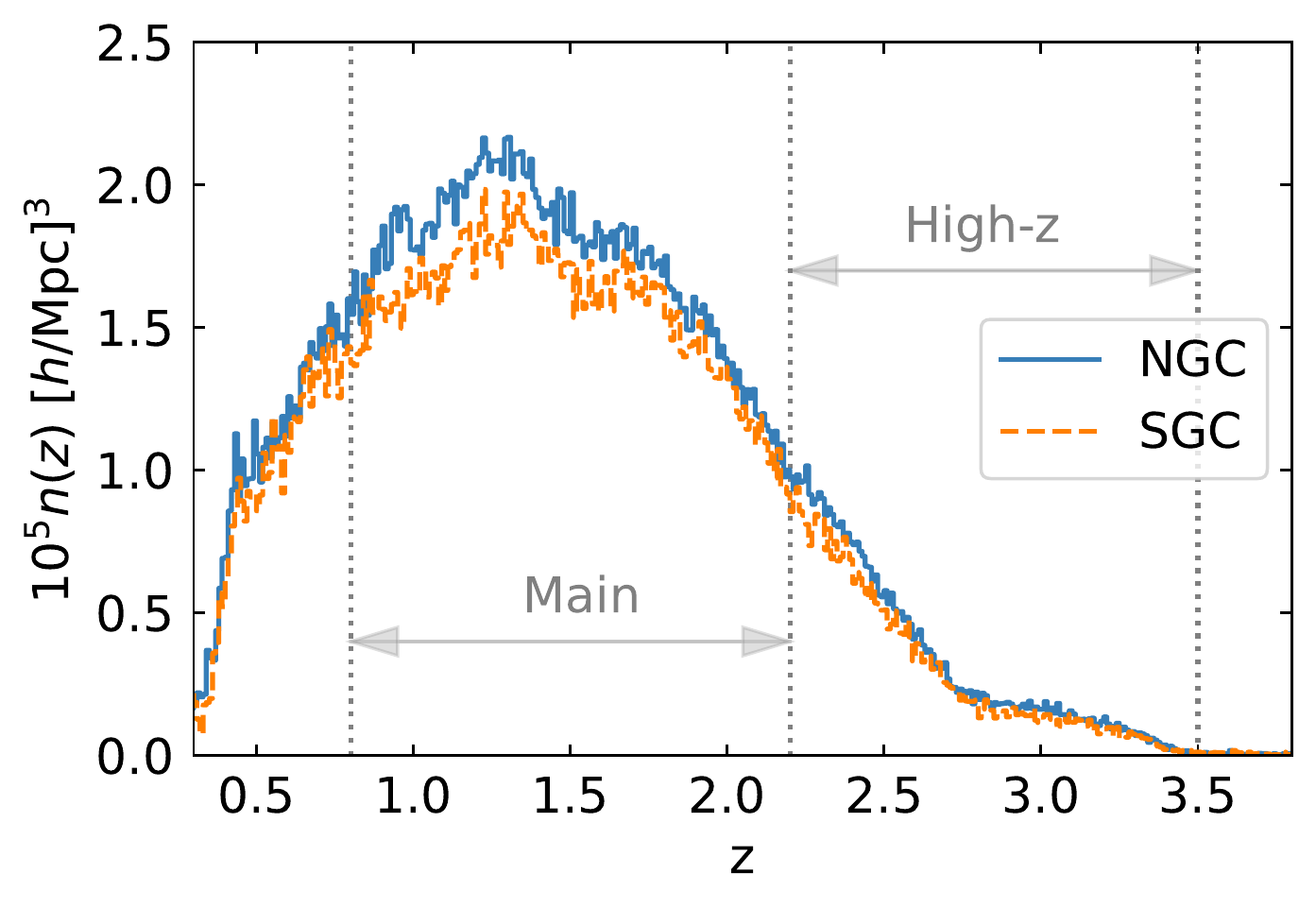}
    \caption[Density of eBOSS DR16 quasars vs redshift.]{Mean density of the DR16 quasars as a function of redshift for the NGC (solid blue) and SGC (dashed orange) regions. The vertical dotted lines represent the redshift cuts for selecting the main and high-z samples.}
    \label{fig:nz}
\end{figure}

\begin{figure}
    \centering
    \includegraphics[width=0.45\textwidth]{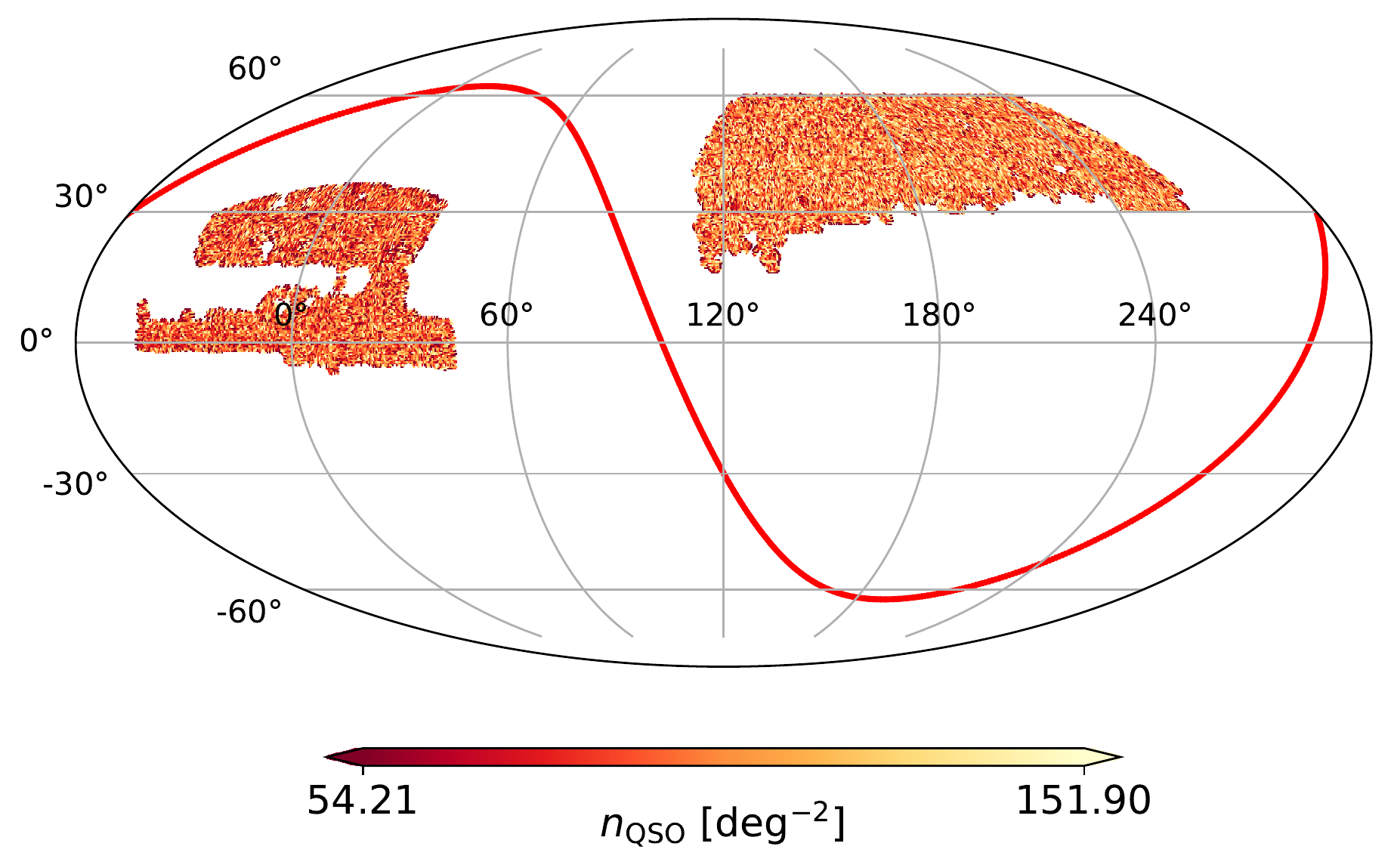}
    \includegraphics[width=0.45\textwidth]{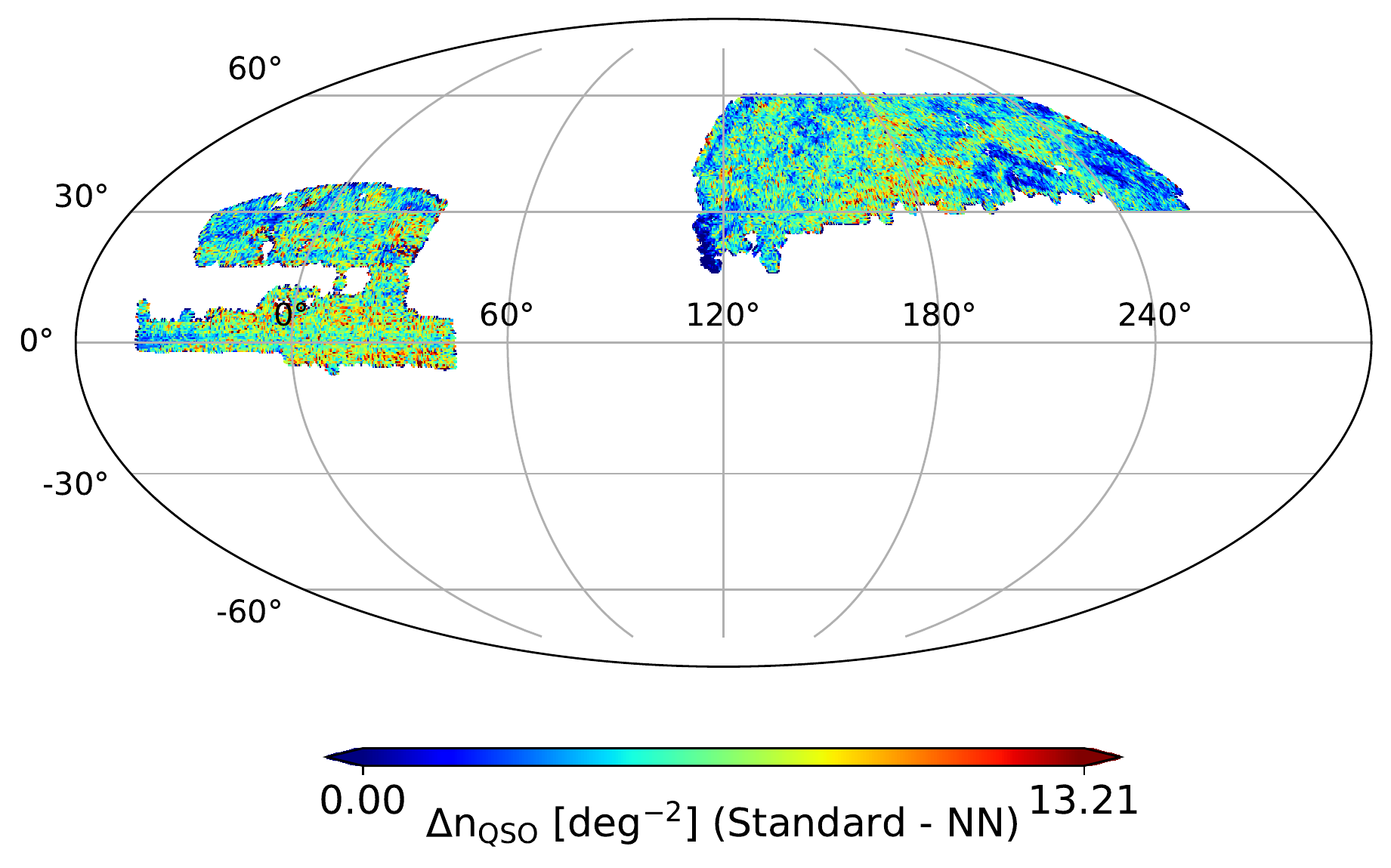}    
    \caption[Mollweide projection of DR16 quasars.]{\textit{Top}: Mollweide projection of the DR16 sample in the redshift interval of $0.8<z<3.5$ and the effective area of $4700~ {\rm deg}^{2}$ with the standard treatment for imaging data systematics. The solid red curve represent the Galactic plane. \textit{Bottom}: The difference in corrected quasar density between applying the standard or neural network treatments for imaging data systematics.}
    \label{fig:mollweide}
\end{figure}

Fig.~\ref{fig:nz} shows the mean number density of quasars from the DR16 catalogue as a function of redshift for the NGC (solid blue) and SGC (dashed orange) regions. The main and high-z samples are separated by vertical dotted lines. Targeting efficiency varies slightly between the two caps which causes small differences in $n(z)$. Figure~\ref{fig:mollweide} shows the sky coverage of the DR16 quasars in equatorial coordinates. The top panel is the Mollweide projection\footnote{The density map is in \textsc{healpix} \citep{gorski2005healpix} with $\textsc{nside}=128$, which corresponds to a pixel area of $0.21~{\rm deg}^{2}$} of the quasar density field in ${\rm deg}^{-2}$ after the standard treatment. The solid red curve represents the Milky Way plane. The effective area of the NGC and SGC regions are $2860$ and $1839~{\rm deg^{2}}$, respectively \citep[see table 3 in][]{ross2020lss}. The quasar density in each pixel is corrected for pixel completeness using the number of random objects. The bottom panel describes the residual quasar density between the standard and new neural network treatments. These methods are thoroughly described in \S \ref{subsec:mitigation}. Interestingly, the residual map indicates that the size of the correction depends on angular direction, with the nonlinear approach leading to more correction in particular regions.

\subsection{Synthetic Catalogues}
\label{subsec:mocks}
We use synthetic catalogues \citep{zhao2020qso}, generated using the extended Zeldovich approximation \citep{zel1970gravitational,chuang2015ezmocks}, to construct covariance matrices of our statistics and perform robustness tests, characterizing the significance of residual systematic uncertainties. Throughout this manuscript, we refer to these mocks as the \textit{EZmocks}. The EZmocks are tuned to reproduce accurate two-point clustering statistics, e.g., within $1\%$ of an N-body simulation on $k<0.55~h/{\rm Mpc}$ and $r > 10~{\rm Mpc}/h$ \citep{chuang2015ezmocks}, and thus are suitable for studying the large-scale clustering of galaxies and quasars. These simulations are sufficient for this work since the impact of imaging systematics on quasar power spectrum measurements appears primarily on long-wavelength modes, e.g., $k < 0.01~h/{\rm Mpc}$. A detailed description for the creation of the eBOSS EZmocks is presented in \cite{zhao2020qso}.

An EZmock realization is created by adopting the Zeldovich approximation to generate a density field. Then, stochastic and parametric techniques are applied to simulate the scale-dependent nonlinear biasing effects. A flat $\Lambda$CDM universe defined by cosmological parameters $h=0.678$, $\Omega_{M}=0.307$, $\Omega_{\Lambda}=0.693$, $\Omega_{b}=0.0482$, $\sigma_{8}=0.822$, and $n_{s}=0.961$ is considered as the input cosmology for the EZmocks. The mock realizations do not contain primordial non-Gaussianity, i.e., $f_{\rm NL}=0$. Each mock catalogue is constructed by combining seven periodic boxes with the comoving side length $L=5~{\rm Gpc/h}$ to mimic a light-cone geometry which accounts for the redshift evolution of quasars. Then, the mock catalogue is sub-sampled along the line of sight to simulate the redshift distributions of the DR16 quasar sample.

We use two sets of simulations in this paper. Each set contains roughly $1000$ independent realizations per Galactic cap.  We have the mocks only for the redshift interval $0.8<z<2.2$. One set of the mocks is manipulated to reflect the effects of observational systematics, including fiber collision, stellar contamination, redshift failures, and angular imaging systematics. While the other set lacks any observational systematics, except the survey geometry effect. Throughout this manuscript, we refer to these two sets as the \textit{contaminated} or \textit{null} mocks, indicating whether or not observational systematic effects are added to the simulations. In particular importance to our work, angular imaging systematics are simulated based on a linear multivariate fit to the DR16 quasar density against two imaging templates\footnote{This is a conservative approach for simulating imaging systematics in the mock catalogues since the treatment of the real sample has shown that more than two templates are required to achieve a desirable cleaning.} for the SDSS stellar density and depth-g with \textsc{healpix} resolution of $\textsc{nside=512}$ (see \ref{subsec:templates}). Similar to the DR16 sample, eBOSS pipelines are applied to the EZmock realizations to calculate and assign the appropriate attributes, which are intended to mitigate the simulated systematic effects and recover the ground truth clustering.

\subsection{Imaging Templates}\label{subsec:templates}

In total we have 17 templates of photometric variables, that are involved in the SDSS imaging and targeting, as potential sources of systematic uncertainties. These templates comprise three maps tracking the structure of the Milky Way: \textit{Galactic extinction} \citep{schlegel1998maps}, \textit{neutral hydrogen column density} \citep{Hi4pi2016}, and \textit{stellar density} from either the SDSS program \citep[as used in][]{bautista2018sdss} or the Gaia spacecraft \citep{gaia2018}. The survey-specific templates are \textit{seeing}, \textit{sky brightness}, and \textit{survey depth} in four bands (\textit{griz}). The remaining two maps are \textit{run} and \textit{airmass}. 

We produce these imaging templates in \textsc{healpix} by using a catalogue of uniformly distributed random objects painted with imaging features. For each pixel, the average of each imaging property is computed over the random objects in the pixel. The maps are created in the ring ordering format and two resolutions, \textsc{nside}=$256$ and $512$, respecively corresponding to $0.23$ and $0.11$ degrees. Although the standard treatment uses templates with \textsc{nside}=$512$ for cleaning the eBOSS DR16 quasars, the preparation of templates in a different resolution, e.g., $256$, will enable us to investigate how changing pixel size might influence the mitigation process and quasar clustering measurements. 

A caveat of all template-based cleaning methods is that the available templates are not perfect, and this limitation might affect the performance of such treatments; for instance, the default stellar density map is an incomplete sample of stars, which is constructed as the number of type PSF objects with $i < 19.9$ from SDSS. Therefore, our proxy of stellar artifacts may be an incomplete representation of existing stellar contaminations in the DR16 sample. Indeed, as we discuss later in \S \ref{sec:results}, we find a noticeable reduction in residual systematic error by swapping the SDSS map for a different stellar map from the Gaia DR2 with $12 < g < 17$ \citep{gaia2018}. The other caveat is that using all imaging maps for training might add the input noise to the neural network, increase chance cross-correlation between cosmological signal and imaging fluctuations, and reduce the effective number of modes.
Therefore, we decide to follow a conservative approach to use a minimum number of templates in regression analysis. Later in \S \ref{sec:results}, our results show that spurious fluctuations against imaging properties are alleviated by using only fives maps including \textit{sky-i}, \textit{seeing-i}, \textit{extinction}, \textit{depth-g}, and \textit{Gaia stellar density}. Based on the $\chi^{2}$ of the mean density residuals, the first four maps are identified as the primary sources of systematic error in the standard cleaning procedure \citep{ross2020lss}, and hereafter we refer to this set as the \textit{known} templates.
\section{Analysis techniques}
\label{sec:techniques}

This section presents the techniques and statistics employed in this work for the characterization and mitigation of residual spurious fluctuations, and outlines the estimator for measuring the large-scale clustering of the DR16 sample and that of the EZmock realizations.

\subsection{Template-based Mitigation}\label{subsec:mitigation}

\subsubsection{Linear Multivariate Regression}\label{subsubsec:linmethod}

The default systematic weights in the DR16 catalogue \citep{ross2020lss} are obtained based on a linear multivariate regression using the four known maps, i.e., \textit{sky-i}, \textit{seeing-i}, \textit{extinction}, and \textit{depth-g}, as the primary sources of systematic fluctuations. Linear regression is the most commonly used approach to reduce the effects of imaging systematics in previous SDSS catalogues \citep[e.g.,][]{ross2011ameliorating, ross2012MNRAS.424..564R, myers2015sdss, prakash2016sdss, ross2017MNRAS.464.1168R}. Despite some minor differences among the various implementations, a common assumption is that the observed number density of quasars in pixel $i$ is a linear combination of imaging quantities in the pixel \citep[][eq. 6]{bautista2018sdss},
\begin{equation}
    y_{i} = \bar{n} +  \sum_{j=1} p_{j} x_{j, i},
\end{equation}
where $\bar{n}$ is the mean density of quasars, $x_{j,i}$ is the $j$'th imaging variable in pixel $i$ and $p_{j}$ is the corresponding linear coefficient. These methods mostly differ either on their cost function\footnote{Cost function is used to find the best parameters of a model.} or on the number of templates used in the model. For instance, the least square error is evaluated over either the binned or pixelated mean density of quasars. Sometimes, the regression routine fits against all templates simultaneously or one template at a time in each iteration. 

Specifically, the method presented in \cite{bautista2018sdss} implements a simultaneous fit using the four known maps as the independent variables in the model and chooses the least square error on the binned quasar density as the cost function. A subtle aspect of the cost function is that the residual fluctuations are summed over not only the four known maps but also the SDSS stellar density and airmass. This design will train the model parameters to explain the trends against the six templates while providing only the four known maps as input. Given the four known maps as the input variable $x$, the systematic weight of quasars in pixel $i$ is defined as, $w_{{\rm systot}, i}=1/(c + \sum_{j}p_{j}x_{j,i})$, where $c$ is the intercept. The optimal coefficients and intercept $c$ are then derived by minimizing the least-squares error\footnote{It is worth noting that unlike the neural network method, this approach uses all data for training and does not apply any form of training, validation, and testing. We believe linear regression is not in a limit prone to over-fitting, i.e., the degree of freedom is much smaller than the number of data points.} on the binned quasar density after applying the systematic weight $w_{\rm systot}$ to each quasar object. The binned density is accounted for other observational and pixel completeness effects by weighting the quasars and randoms respectively by $w_{\rm tot, g}$ and $w_{\rm tot, r}$, see, Eq. \ref{eq:wbefore_after}.

\subsubsection{Non-Linear approximation using Neural Networks}\label{subsubsec:nnmethod}

The DR16 quasar sample is very sparse, with a surface density of $73~{\rm deg}^{-2}$ in the redshift range $0.8<z<2.2$ (main) and $15~{\rm deg}^{-2}$ in $2.2<z<3.5$ (high-z). The high sparsity poses a major challenge for regression-based treatments by making it difficult to disentangle the effect of systematics and the cosmological signal. This paper expands the neural network-based methodology described in \cite{rezaie2020MNRAS.495.1613R} to obtain nonlinear systematic weights. In what follows,  we first describe the specifics of the neural network architecture, and then elaborate on how we further improve the methodology specifically with the implementation of the Poisson statistics in the cost function to properly deal with the high sparsity eBOSS QSO data and a cyclic learning rate to enhance training against local minima.

\textbf{Feed forward neural networks:} Neural networks are universal approximators that can model a wide variety of non-linear mappings from a set of independent variables to a target variable. Neural networks can account for nonlinear spurious fluctuations caused by imaging and the cross-correlation among imaging properties. We use $y_{i}$ and $\textbf{x}_{i}$ to respectively denote the observed quasar count and imaging features\footnote{We use the z-score normalization scheme to standardize our imaging templates, i.e., $z = (x - \mu)/\sigma$ where $\mu$ and $\sigma$ are the mean and standard deviation of imaging feature $x$ determined from the training set.} in pixel $i$. We implement a fully connected feed-forward neural network to model the observed quasar count $y_{i}$ from imaging templates as input features $\textbf{x}_{i}$. At its standard configuration, a fully connected feed-forward neural network is a system of neurons organized in a series of interconnected layers, where each neuron is connected to all neurons in the previous layer and the following layer\footnote{Bias neurons are an exception and connect only to the subsequent layer.}. For instance, the output value of neuron $\mu$ in layer $l+1$, $a_{\mu}^{l+1}$, is obtained from the linear combination of the output values from the previous layer neurons, $a_{\nu}^{l}$, after applying a nonlinear activation function $f$,
\begin{equation}
    a_{\mu}^{l+1} = \psi(b^{l}_{\mu} + \sum_{\nu} w^{l}_{\mu \nu}a_{\nu}^{l}),
\end{equation}
where $w^{l}_{\mu \nu}$ and $b^{l}_{\mu}$ are respectively the associated weights and bias connecting to neuron $\mu$. The nonlinear function $\psi$ determines how much of the signal travels from layer $l$ to $l+1$. In this notation, the first layer input values are imaging properties $x_{i}$ and the last layer output is then compared to $y_{i}$. The network architecture comprises three fully-connected hidden layers\footnote{We follow a grid search approach to experiment with a various number of hidden layers ranging from two up to five hidden layers, however we do not observe any significant change in validation loss. Therefore, we fix the architecture at three hidden layers with 20 units on each hidden layer.} with twenty neurons with the rectified linear activation function, $\psi(u)=\max(0, u)$, on each hidden layer and a single neuron with the softplus activation function, ${\rm Softplus}(u)=\log[1+\exp(u)]$, on the output layer. The softplus activation in the last layer will ensure the network output is always positive. We also add a batch normalization layer after each layer, except the output layer. Batch normalization stabilizes and expedites training by scaling the inputs to each layer to have a mean of zero and a standard deviation of one \citep[e.g.,][]{Ioffe2015}.

{\bf Poisson statistics in the cost function:} For the DR16 quasars, we decide to use the negative Poisson log-likelihood as the default cost function and the Mean Squared Error for benchmarking. Assuming the quasar counts per pixel are independent and identically distributed variables, we can write the joint probability for $N$ pixels as,
\begin{equation}
     L = f(y_{1}, ..., y_{N} | \theta) = \prod_{i=1}^{N} f(y_{i}|\theta, \textbf{x}_{i}),
\end{equation}
which is a reasonable assumption provided that the input features $\textbf{x}_{i}$ do not include any spatial information, and we do not intend to provide any feature that contains cosmological clustering in our modeling. Otherwise, the mitigation procedure will learn the clustering signal and remove it as well. The objective is then to find the best set of parameters $\theta$ that maximizes the likelihood $L$, or minimizes its negative logarithm $-\log L $. Assuming $y_{i}$ follows a Poisson distribution, we have
\begin{equation}
    f(y_{i}|\theta,x_{i}) = \frac{\lambda(\theta, \textbf{x}_{i})^{y_{i}} \exp^{-\lambda(\theta, \textbf{x}_{i})}}{y_{i}!},
\end{equation}
where $\lambda$ is the expected number of quasars, $\lambda>0$ by definition, and a function of imaging features. We use $\lambda_{i}\equiv\lambda(\theta, \textbf{x}_{i})$ for brevity and obtain the negative log likelihood\footnote{We omit the term $\log(y_{i}!)$ from our objective function since it does not depend on $\theta$.} as the cost function $J$,
\begin{align}\label{eq:loglik}
    J \equiv -\log (L) = \sum_{i=1}^{N} [\lambda_{i} - y_{i} \log(\lambda_{i})].
\end{align}
In this notation, the quasar number count $y_{i}$ is assumed to be an integer value. This assumption breaks down in practice since each object in the DR16 catalogue needs to be weighted for other observational systematics and pixel completeness beforehand, and that turns $y_{i}$ into a non-integer value. To circumvent this issue and still account for the pixel completeness and other observational effects, we decide to use the raw number count of quasars for $y_{i}$, and instead weight the model prediction in pixel $i$, i.e., $\lambda_{i}$, by the ratio of the weighted number of randoms to that of quasars in that pixel. To this end, the quasar and random objects are respectively weighted by $w_{\rm tot, g}$ and $w_{\rm tot, r}$,
\begin{align}\label{eq:wbefore_after}
    w_{\rm tot, g} &= w_{\rm noz} \times w_{\rm cp} \times w_{\rm FKP}, \nonumber\\
    w_{\rm tot, r} &= w_{\rm FKP} \times {\rm comp}_{\rm BOSS},
\end{align} 
where ${\rm comp}_{\rm BOSS}$ is the survey completeness that describes the likelihood of obtaining a good redshift, regardless of any other instrumental deficiencies.

{\bf Cyclic learning rate:} The weights $w^{l}_{\mu \nu}$ and biases $b^{l}_{\mu}$ of the neural network are trained using the Decoupled Weight Decay Regularization optimizer \citep[AdamW;][]{LoshchilovAdamW}, which is an iterative gradient descent approach for optimization. Specifically, the parameters are updated in the opposite direction of the gradient of the cost function with respect to the parameters,
\begin{align}
    m_{t+1} &= \beta_{1}m_{t} + (1-\beta_{1}) \nabla J(\theta_{t}), \\
    v_{t+1} &= \beta_{2}v_{t} + (1-\beta_{2}) [\nabla J(\theta_{t})]^{2}, \\
    \theta_{t+1} &= \theta_{t} - \eta_{t} \frac{m_{t+1}}{\sqrt{v_{t+1}}+\epsilon},
\end{align}
where $\epsilon=10^{-8}$ and the learning rate $\eta_{t}$ controls the magnitude of each parameter update per iteration $t$. The first and second moments of the gradients, i.e., $m_{t}$ and $v_{t}$, are initialized as zero. The parameters $\beta_{1}$ and $\beta_{2}$ determine the average history of the first and second moments of the gradients, and are fixed at $0.9$ and $0.999$, respectively \citep[see, e.g.,][for a review of gradient descent methods]{ruder2016arXiv160904747R}.

We incorporate a cyclic learning rate to prevent the optimizer from being trapped in a local minimum. Specifically, the learning rate $\eta$ at epoch $t$ is scheduled to vary following the method presented in \citep{Loshchilov2016},
\begin{equation}
        \eta_{t} = \eta_{\rm min} + \frac{1}{2}(\eta_{\rm max} - \eta_{\rm min}) [\cos(\frac{T_{\rm cur}}{T_{i}}\pi)+1],
\end{equation}
where $\eta_{\rm max}$ is the initial (maximum) learning rate, $\eta_{\rm min}$ is the minimum learning rate, $T_{\rm cur}$ is the number of epochs since the last restart, and $T_{i}$ is the number of epochs between two subsequent restarts. The neural network training begins with $T_{i=0}=10$, but then we increase $T_{i}$ by a factor of two after each restart (i.e., $T_{i}$ changes like 10, 20, 40, ...). We follow the procedure presented in \cite{Smith2015} to search an exponential grid\footnote{\href{https://github.com/davidtvs/pytorch-lr-finder}{github.com/davidtvs/pytorch-lr-finder}; \href{https://github.com/fastai/fastai}{github.com/fastai/fastai}} for the optimal values of $\eta_{\rm min}$ and $\eta_{\rm max}$. The optimal values are chosen such that the variation of loss per training epoch is maximized. We separately perform the learning rate finding procedure for each Galactic cap, redshift selection, template resolution since the balance between systematics and noise changes.

\begin{figure}
    \centering
    \includegraphics[width=0.45\textwidth]{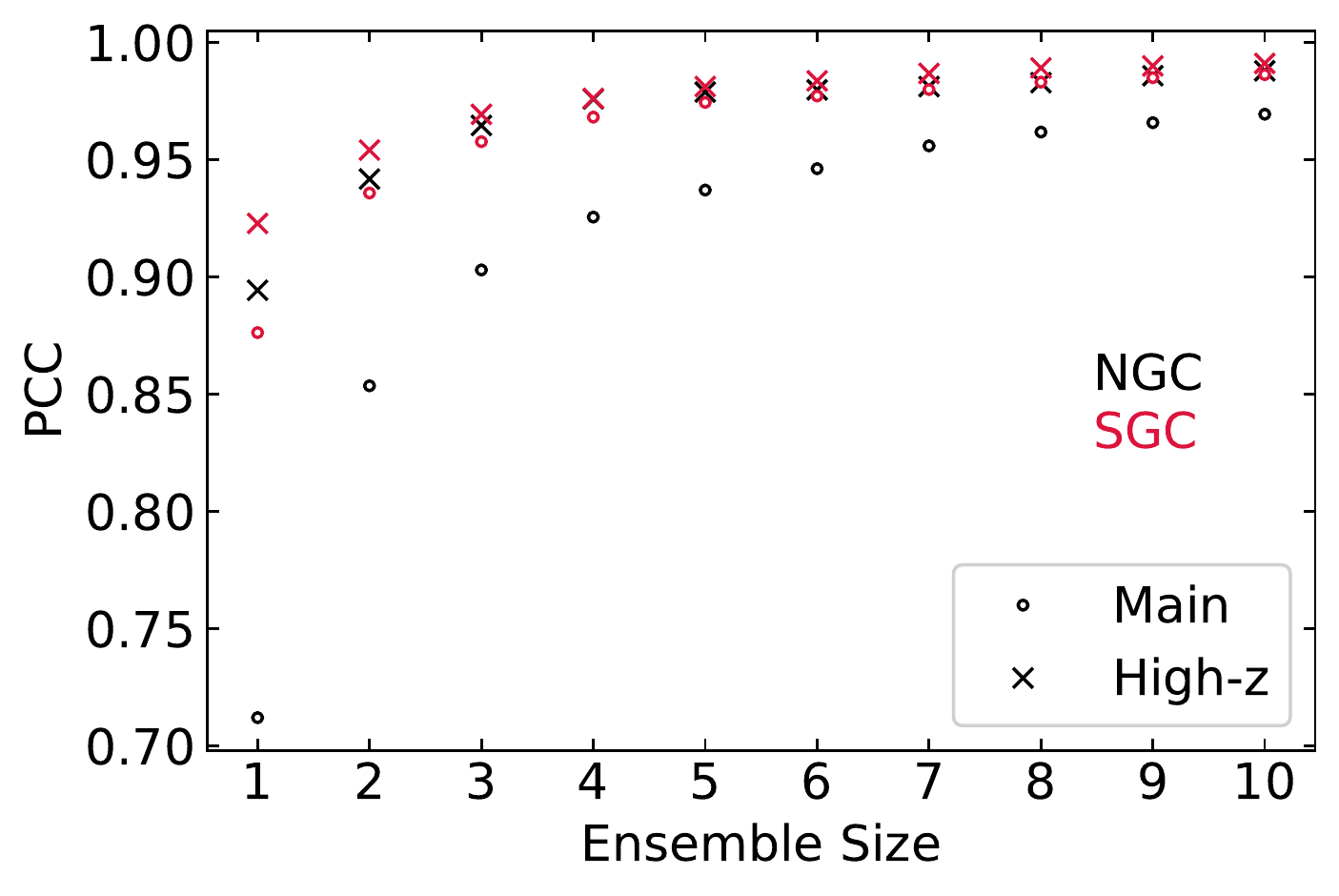}
    \caption[Pearson's correlation between predicted quasar counts vs ensemble size.]{The Pearson correlation coefficient between the averaged predicted quasar counts from neural networks for various ensemble size in the NGC (black) and SGC (red) regions, and the main and high-z samples. The correlation increases beyond $90$ per cent after averaging over three networks.}
    \label{fig:pccnpred}
\end{figure}

{\bf Cross-validation:} We apply five-fold cross-validation to perform training, validation, and testing the network on the entire footprint, while ensuring no overlap among any of the training, validation, or testing sets. Specifically, we randomly split the entire footprint into $60\%$ training, $20\%$ validation, and $20\%$ testing sets. The training set is used to compute the gradients and update the parameters. In each epoch, the model is applied to the validation set to assess the prediction error when the model is applied on unseen data. We let the networks to train for $150$ epochs, and finally we apply the model with the lowest validation error on the test set. By changing the arrangement of the training and validation sets, the neural network will be tested on the entire footprint. Similar to \cite{ross2020lss}, we perform training, validation, and testing for the main and high-z samples in the NGC and SGC regions separately and finally aggregate the results. This approach assumes that the effects of imaging systematics along the line-of-sight do not vary much. We relax this assumption later by further splitting the main sample into $0.8<z<1.5$ and $1.5<z<2.2$, and analyzing each subsample separately. As shown later in \S \ref{sec:results}, we find no significant difference. Therefore, we leave an extensive 3D regression of target density and imaging properties for future work. 

Unlike linear regression, the objective function of a neural network is non-convex, and thus there is no unique set of best-fit parameters. This results in an intrinsic scatter on the predicted quasar density from a single neural network. We create an ensemble of 20 networks, through random initializations of the parameters, to mitigate the issue of model uncertainty, reduce the dispersion of the predicted quasar counts, and increase the stability of the selection function. We then take the mean predicted number of quasars across all networks in the ensemble as the final selection function. Fig. \ref{fig:pccnpred} illustrates the Pearson correlation coefficient (PCC\footnote{PCC$(X, Y) = cov(X, Y) / \sqrt{cov(X,X)cov(Y, Y)}$, where $cov(X, Y)$ is the covariance between variable $X$ and variable $Y$.}) between the predicted quasar counts of two independent ensemble subsets as a function of the ensemble size for the main and high-z samples in the NGC (black) and SGC (red) regions. The Pearson coefficient illustrates that the mean predicted densities between two ensembles, each contains only four or more networks, correlate more than $90$ per cent. Then, the systematic weight for quasars in pixel $i$ is defined by $w_{\rm systot} \propto 1/\lambda(\theta, \textbf{x}_{i})$, where $w_{\rm systot}$ is normalized such that the total weighted number of quasars stays the same as before treatment \footnote{We also limit the systematic weights to $0.5 < w_{\rm systot} < 2.0$ to avoid extreme corrections. For pixels where we do not have imaging information, we use the mean value over nearest neighbors to estimate the systematic weight. We do not observe a significant impact on quasar clustering.}. The figure shows that with our default ensemble size of 20, we are introducing very little intrinsic scatter in the process of the neural network.

\subsection{Residual Systematics Tests}\label{subsec:residualtest}

\subsubsection{Mean Density Contrast}\label{subsubsec:meandensity}
This test is sensitive to the uniformity of the DR16 sample by clustering pixels with similar imaging properties. We compute the mean density contrast of quasars against each imaging variable to quantify the deviations and fluctuations in the sample before and after cleaning. For a given imaging quantity $x_{j}$, the mean density contrast $\delta$ is,
\begin{equation}
    \delta(x_{j}) = \frac{\sum_{i} n_{g,i}}{\alpha \sum_{i} n_{r,i}} - 1,
\end{equation}
where $n_{g,i}$ and $n_{r,i}$ are the weighted number of quasars and randoms in pixel $i$; $\alpha$ is the factor to normalize the number of randoms to that of quasars; and the summation $\sum_{i}$ is computed over pixels with $x_{j,i} \in [x_{j}, x_{j}+\Delta x_{j}]$. We adopt an equal frequency binning scheme in which pixels are initially sorted from the minimum to maximum imaging value, and then the width $\Delta x_{j}$ is varied such that each bin contains the same area, i.e., the same number of randoms. By design, this statistic is not sensitive to the underlying true power spectrum, since the mean of density contrast is computed over a large number of pixels in the imaging phase space. Therefore, the cosmological component in $\delta$ cancels out.

\begin{figure}
    \centering
    \includegraphics[width=0.45\textwidth]{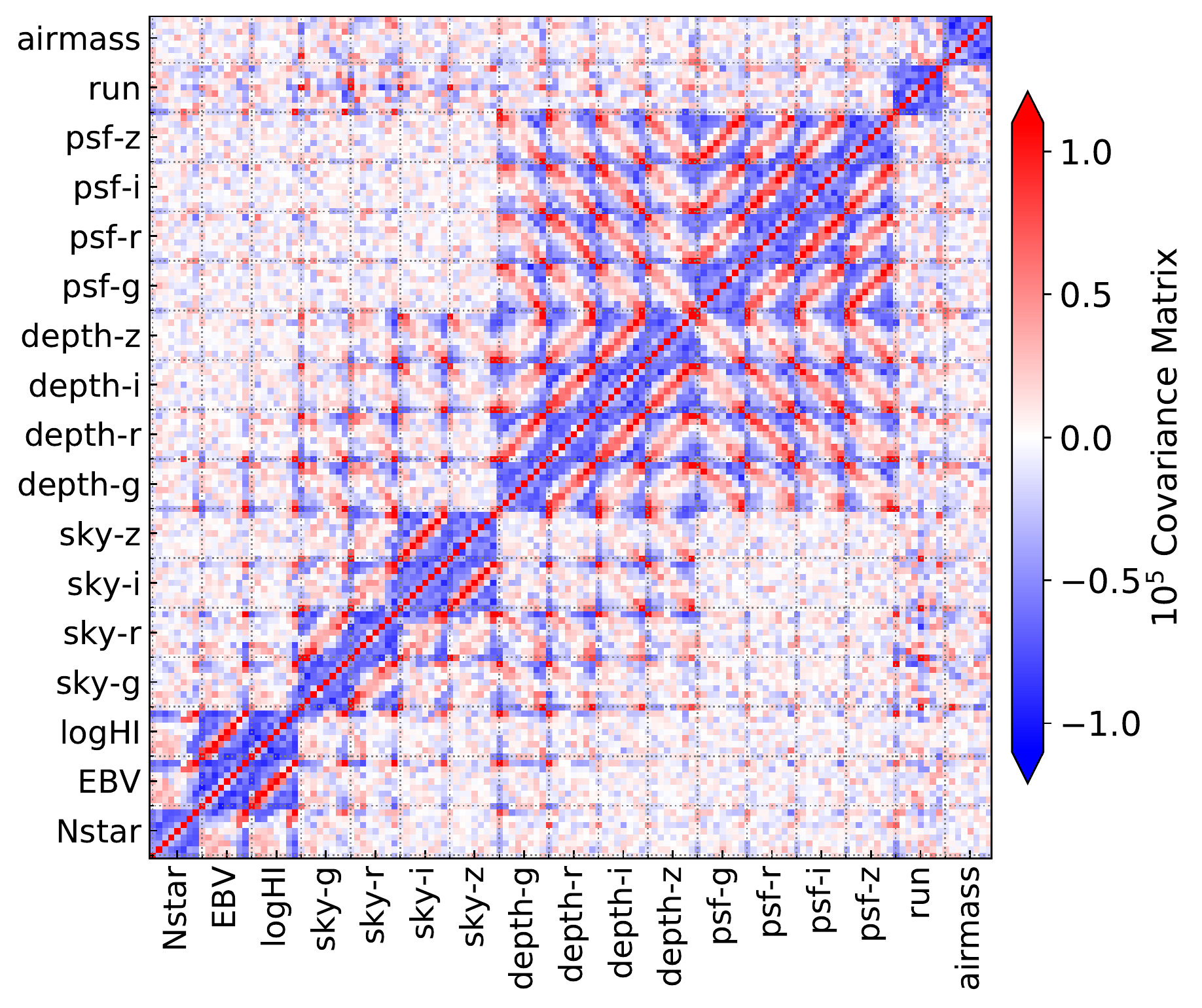}
    \caption[Covariance of mean density contrast]{Covariance matrix of the mean density contrast for the NGC region estimated from the null EZmock realizations.}
    \label{fig:nbarcovmax}
\end{figure}

The quasar and random objects are weighted differently before and after mitigation; initially, the quasars are weighted by $w_{\rm tot, g}$ and the randoms are weighted by $w_{\rm tot, r}$,  Eq. \ref{eq:wbefore_after}. After mitigation, we use Eq. \ref{eq:wtot} to weight both the quasars and randoms. Finally, we compute the mean density contrast against all 17 templates, $\delta = [\delta(x_{1}), \delta(x_{2}),..., \delta(x_{17})]$, and quantify the total mean density residuals with $\chi^{2}$ statistics,
\begin{equation}\label{eq:chi2nbar}
    \chi^{2}_{\rm tot} = \delta^{\dagger} C^{-1} \delta,
\end{equation}
where the covariance matrix $C$ is estimated from the null EZmock mean density contrasts, $C = <\delta^{\dagger}\delta>$, where the angle brackets represent ensemble average over the EZmock realizations. The estimated covariance matrix is then unbiased by the Hartlap factor \citep{hartlap2007A&A...464..399H},
\begin{equation}
    C = \frac{N_{\rm mocks} - 1}{N_{\rm mocks} - N_{\rm bins} - 2}~C_{\rm biased},
\end{equation}
where $N_{\rm mocks}$ is the number of mock realizations, e.g., $1000$ for the NGC and $999$ for the SGC, and $N_{\rm bins}$ is the number of imaging bins, i.e., $136$. Fig.~\ref{fig:nbarcovmax} shows the estimated covariance matrix of the mean density contrast from the $1000$ null EZmock realizations for the NGC. This matrix shows the complex correlation among the attributes. This covariance matrix will be used to assign the error-bars presented in Fig. \ref{fig:data_nbar} and obtain the $\chi^{2}$ values presented in Fig. \ref{fig:chi2_nbar}

\subsubsection{Angular Cross Correlation}\label{subsub:crosscorr}

The anisotropies in target quasar density of a pure cosmological signal should not correlate with Galactic foregrounds and survey-related properties. To test this assumption, we measure the cross power spectrum between quasar density and imaging templates. Due to imaging systematics, the cross power spectrum measurements are not zero, and we use this tool to quantify spurious fluctuations. The cross-spectrum is squared and then normalized by the auto power spectrum of the imaging map,
\begin{equation}
    \Tilde{C}^{q,x}_{\ell} = (C^{q,x}_{\ell})^{2} / C^{x,x}_{\ell},
\end{equation}
to convert it to the estimated contribution of systematics up to the first order to the auto power spectrum of the quasar density. The auto- or cross-power spectrum is computed over the coefficients $a_{\ell m}$ from spherical harmonic decomposition \citep[e.g.,][]{hobson1931theory},
\begin{equation}
    C^{p, q}_{\ell} = \frac{1}{2\ell +1}\displaystyle\sum_{m=-\ell}^{\ell}a^{p}_{\ell m}a^{q}_{\ell m},
\end{equation}
with $p=q$ for auto-correlation and $p\ne q$ for cross-correlation. We bin the cross- and auto-correlations, and use only the four lowest $\ell$ bins centered at $\ell = 1.6,  3.7,  8.3, {\rm and}~18.8$, as we are only interested in large-scale modes\footnote{We use the approximation $\ell+1/2 \sim k D_{A}$ with $D_{A}$ being the angular size distance.}  around $k < 0.01~h/{\rm Mpc}$. Similar to the mean density contrast diagnostic, the covariance matrix $C$ is constructed from the null EZmock cross power spectra, and then unbiased by the Hartlap factor. Finally, we quantify the significance of the residual cross power against the $17$ imaging maps by,
\begin{equation}
    \chi^{2}_{\rm tot} = \Tilde{C}^{\dagger}_{\ell} C^{-1} \Tilde{C}_{\ell},
\end{equation}
where,
\begin{equation}
    \Tilde{C}_{\ell}=[\Tilde{C}^{q,x_{1}}_{\ell}, \Tilde{C}^{q,x_{2}}_{\ell},..., \Tilde{C}^{q,x_{17}}_{\ell}].
\end{equation}

\subsection{Power Spectrum Multipoles}\label{subsec:powerspec}

We use the estimator presented in \cite{hand2017JCAP...07..002H} and implemented in \textsc{nbodykit}\footnote{\href{https://nbodykit.readthedocs.io/en/latest/}{https://nbodykit.readthedocs.io}} \citep{hand2017nbodykit} to measure the power spectrum of the DR16 sample for the NGC and SGC regions, separately. We provide a summary of the algorithm below.

We assume a flat $\Lambda$CDM cosmology with $\Omega_{M}=0.31,h=0.676, \Omega_{b}h^{2}=0.022, \sigma_{8}=0.8, {\rm and}~n_{s}=0.97$ \citep{alam2017clustering} to convert each redshift to distance. After transforming the coordinates, we first begin with digitizing the quasar and random catalogues over a 3D cubic box with a side length of $6600~{\rm Mpc}/h$ and $512^{3}$ voxels. Then, we construct the Feldman-Kaiser-Peacock field \citep[FKP;][]{feldman1994ApJ...426...23F} and interpolate it using the Triangular Shaped Cloud (TSC) scheme,
\begin{equation}
    F(\textbf{r}) = n_{g}(\textbf{r}) - \alpha n_{r}(\textbf{r}),
\end{equation}
where $\alpha=\sum n_{g}(\textbf{r})/\sum n_{r}(\textbf{r})$. The terms $n_{g}$ and $n_{r}$ represent the observed density of the quasar sample and random objects weighted by Eq. \ref{eq:wtot}, respectively. The power spectrum multipoles are then computed from the FKP field as \citep{yamaoto2006PASJ...58...93Y},
\begin{equation}\label{eq:pkestimator}
P_{\ell}(k) = 
\frac{2\ell+1}{4\pi A} \iiint d\Omega d\textbf{r}_{1} d\textbf{r}_{2} F(\textbf{r}_{1}) F(\textbf{r}_{2}) e^{-i\textbf{k}.(\textbf{r}_{2}-\textbf{r}_{1)}} \mathcal{L}_{\ell}(\hat{\textbf{k}.\textbf{r}_{h}})
\end{equation} 
where $\textbf{r}_{h}=(\textbf{r}_{1} + \textbf{r}_{2})/2$ is the line-of-sight distance to the middle point of a given pair, $\mathcal{L}$ is the Legendre polynomial of order $\ell$, and $A=\int d\textbf{r} \bar{n}^{2}(\textbf{r})$ is the normalization for the field $F$, which can be approximated by a discrete sum over the synthetic objects weighted by $w_{r}$ (see, Eq. \ref{eq:wtot}),
\begin{equation}
    A = \alpha \sum_{i} w_{r}(\textbf{r}_{i}) n_{r}(\textbf{r}_{i}).
\end{equation}

Equation \ref{eq:pkestimator} can be simplified even further. Under the small angle approximation, $\hat{\textbf{k}}.\hat{\textbf{r}}_{h}\sim \hat{\textbf{k}}.\hat{\textbf{r}}_{2}$, the two integrals on $r_{1}$ and $r_{2}$ are separable. Using the decomposition of the Legendre function into spherical harmonics,
\begin{equation}
    \mathcal{L}_{\ell}(\hat{\textbf{k}}.\hat{\textbf{r}}) = \frac{4\pi}{2\ell + 1} \sum_{m=-\ell}^{\ell} Y_{\ell m}(\hat{\textbf{k}})Y_{\ell m}^{*}(\hat{\textbf{r}}),    
\end{equation}
we obtain,
\begin{equation}\label{eq:pk}
    P_{\ell}(k) = \frac{2\ell + 1}{A} \int \frac{d\Omega}{4\pi} F_{0}(\textbf{k})F_{\ell}(-\textbf{k}),
\end{equation}
where 
\begin{align}
    F_{\ell}(\textbf{k}) &= \int d\textbf{r} F(\textbf{r}) e^{i\textbf{k}.\textbf{r}}\mathcal{L}_{\ell}(\hat{\textbf{k}}.\hat{\textbf{r}}) \\
    &= \frac{4\pi}{2\ell +1}\sum_{m=-\ell}^{\ell} Y_{\ell m}(\hat{\textbf{k}}) \int d\textbf{r} F(\textbf{r})e^{i\textbf{k}.\textbf{r}} Y_{\ell m}^{*}(\hat{\textbf{r}}).
\end{align}
We measure the power spectrum using Eq. \ref{eq:pk} in k bands of width $\Delta k=0.0019~ h{\rm Mpc}^{-1}$. This method requires less number of Fourier's transforms compared to other estimators in \cite{bianchi2015MNRAS.453L..11B, scoccimarro2015PhRvD..92h3532S}. When computing the transforms, the \textsc{nbdoykit} software mitigates aliasing with the interlace grid technique presented in \cite{Sefusatti2016MNRAS} and accounts for any effects caused by the TSC gridding using the factor presented in \cite{jing2005ApJ}. Finally, we estimate the shotnoise by summing over the quasar and random catalogue objects, 
\begin{equation}\label{eq:shotnoise}
    P_{\rm shotnoise} = \frac{1}{A}[\sum_{i} w^{2}_{g}(\textbf{r}_{i}) + \alpha^{2} \sum_{j} w^{2}_{r}(\textbf{r}_{j}) ],
\end{equation}
and subtract it only from the monopole power spectrum ($\ell=0$).
\section{Results}\label{sec:results}

This section presents the analyses for the characterization of systematic error and the treatment of spurious fluctuations with various cleaning approaches. We compare the measured power spectra before and after accounting for imaging systematic effects with various linear and nonlinear cleaning methods. In the end, we present the clustering analysis of the EZmock realizations in order to assess the impact of systematic treatments on the measured power spectrum and how one can account for such effect in theoretical modeling.

\subsection{Method Benchmarks}

\subsubsection{Linear vs Nonlinear Treatment}\label{subsec:linvsnonlin}

Our series of tests begin with applying various linear and nonlinear cleaning methods to the main sample of quasars in the NGC region. To conduct a fair comparison, the same set of the four known maps are employed as the input templates. We use the Mean Squared Error and the Poisson Negative Log-Likelihood as two alternatives for the cost function. When using PNLL as the cost function, \textit{Softplus} is applied to the output to satisfy the boundary condition $\lambda > 0$. We also experiment with a nonlinear model without the learning rate annealing to test the sensitivity of training to local minima.

\begin{figure}
    \centering
    \includegraphics[width=0.45\textwidth]{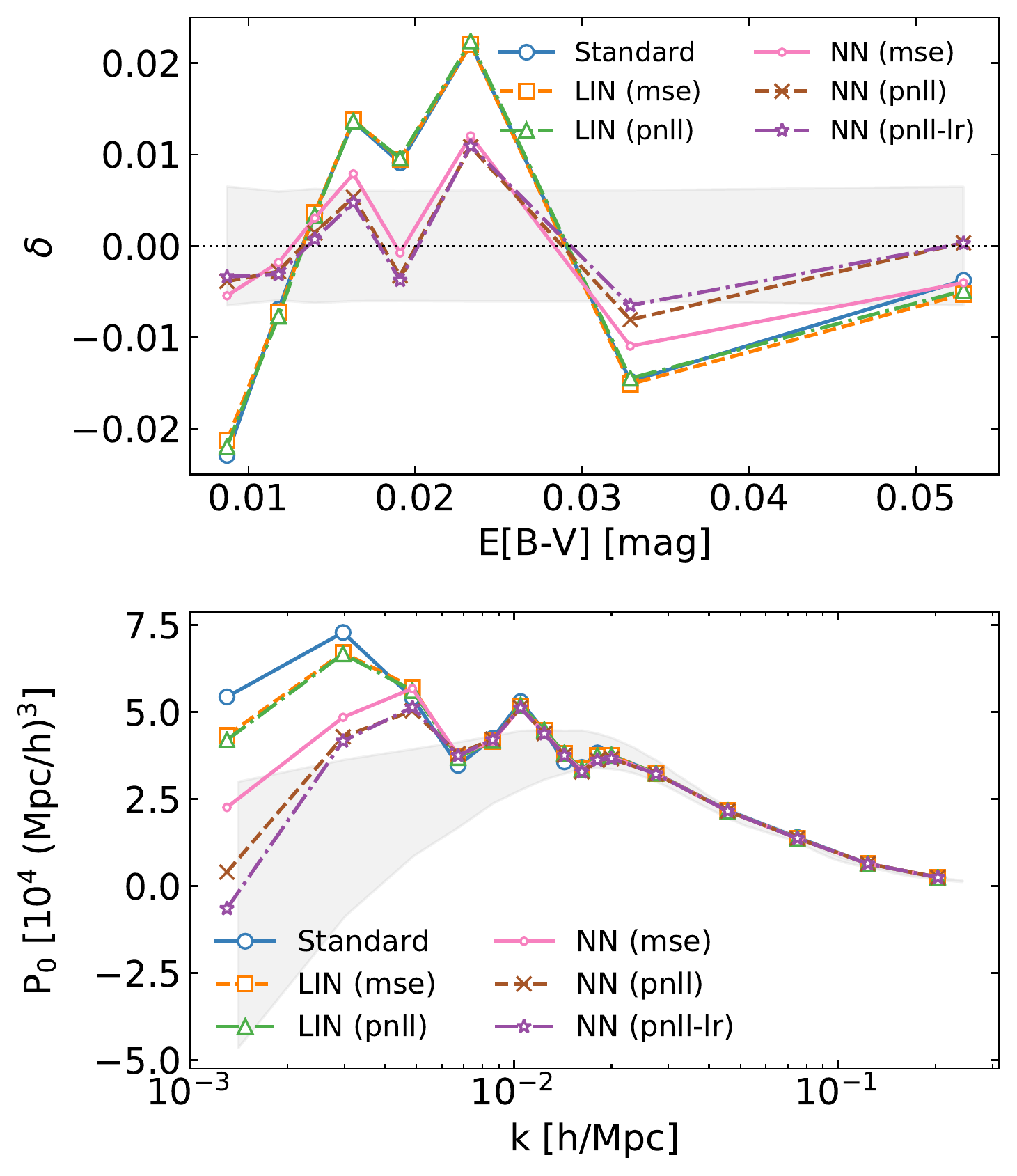} 
    \caption[Mean density and power spectrum of eBOSS quasars for different methods.]{\textit{Top}: Mean density contrast of the NGC main sample against Galactic extinction after applying various mitigation techniques, including linear with MSE, linear with PNLL, NN with MSE, NN with PNLL, and NN with PNLL but without learning rate annealing (NN-PNLL-lr). All methods use the \textit{known} set of imaging maps.  \textit{Bottom}: Monopole of the NGC main sample after applying the same techniques.  The shades represent 1$\sigma$ statistical uncertainty constructed from the null EZmock realizations.}
    \label{fig:nbarmethods}
\end{figure}

Fig. \ref{fig:nbarmethods} shows the mean density contrast as a function of Galactic extinction in the top panel and the measured power spectrum in the bottom panel, after applying different cleaning methods. The 1$\sigma$ uncertainty is shown via the grey shades and is constructed from the null EZmock realizations. Linear-MSE here is  equivalent to the standard treatment except that the cost function of the former is based on the pixelated density while the latter is based on the binned density. This figure illustrates that our implementations of linear-MSE and linear-PNLL yield similar residual fluctuations to that of the standard treatment. This result implies that with linear regression, adopting the Poisson statistics does not help. Interestingly, the residual fluctuations are reduced substantially after accounting for non-linearities by the hidden layers in the \textit{NN-MSE} and \textit{NN-PNLL} architectures. We also find that turning off the learning rate cycling does not hugely impact the neural network performance (\textit{NN-PNLL-lr}) with marginal effect on the power spectrum, which proves the robustness of the pipeline against local minima and saddle points. This test shows that the most substantial improvement in the measured power spectrum is enabled by accounting for nonlinear systematic effects using the neural network-based methods and then by adopting PNLL in the cost function. Interestingly, our linear cleaning method yields a lower clustering power than the standard linear approach. This test demonstrates that cost function needs to be defined accurately for different levels of model complexity; the choice of PNLL over MSE makes more substantial difference for the NN approach compared with the linear model.

\subsubsection{Stellar Density from Gaia DR2}\label{subsec:nstar}
The quality of the selection function derived from a template-based cleaning method relies on the available input templates. We may fail to properly eliminate spurious fluctuations if a primary imaging map is missed or our available imaging maps are not completely representing the underlying systematic effects. In this test, we experiment with the available SDSS and non-SDSS imaging maps to find the optimal number of crucial imaging maps, which one would need to explain the residual trends in the mean density contrast. As mentioned before, the standard cleaning approach uses only four maps, i.e., Galactic extinction, depth in g band, sky brightness in i band, and seeing in i band.

\begin{figure}
    \centering
    \includegraphics[width=0.45\textwidth]{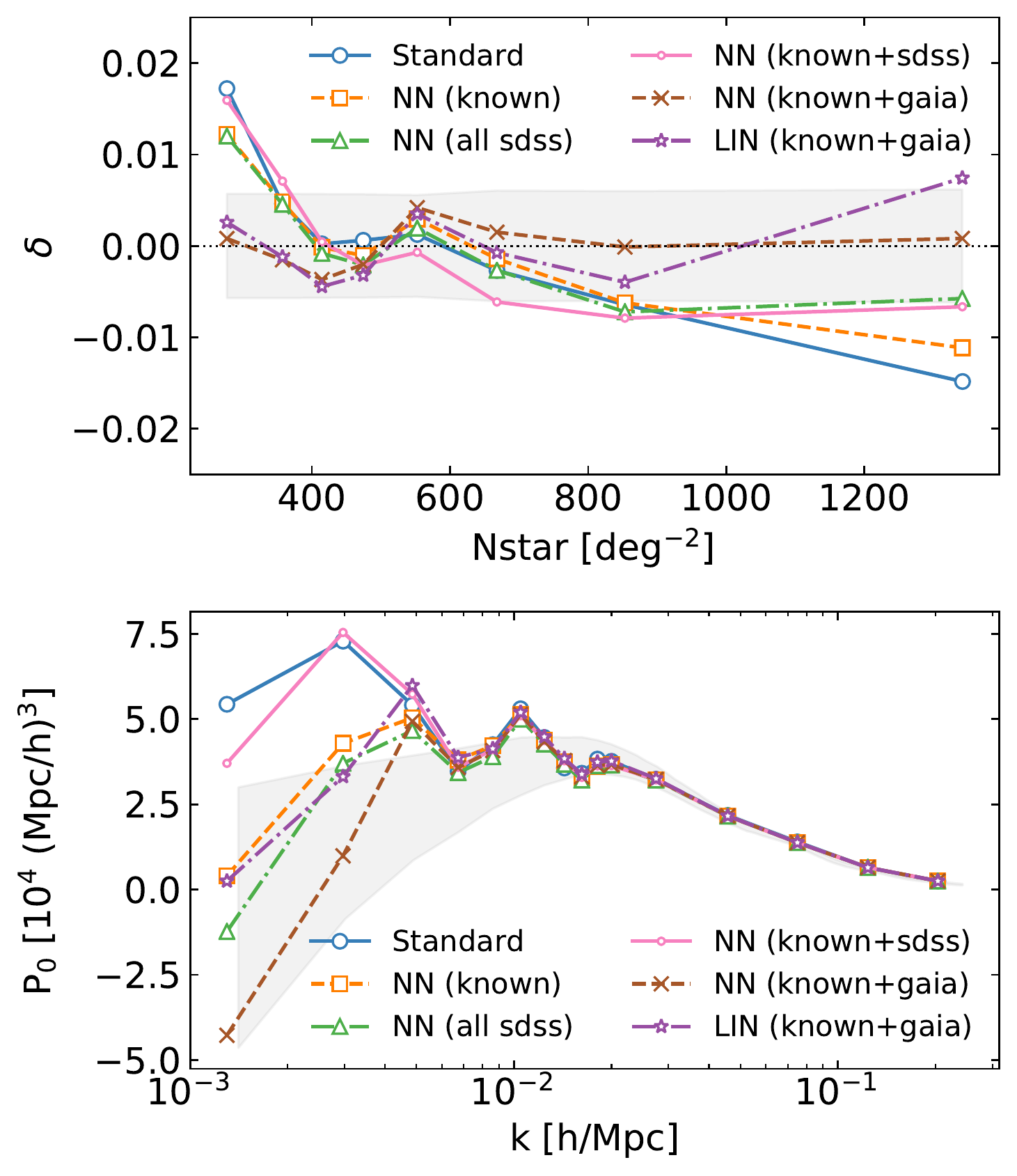}
    \caption[Mean density and power spectrum of eBOSS quasars for different templates.]{\textit{Top}: Mean density contrast of the NGC main sample against the Gaia stellar density after accounting for systematics using different combinations of imaging templates \textit{Bottom}: Monopole power spectrum of the NGC main sample for the same techniques. The shades represent 1$\sigma$ statistical uncertainty constructed from the null EZmock realizations.}
    \label{fig:nbarnstar}
\end{figure}

Fig. \ref{fig:nbarnstar} top panel shows the mean density contrast against the Gaia stellar density from \cite{gaia2018} after training a neural network with various combinations of the imaging maps using PNLL as the cost function. We also plot the measured density contrast after the standard treatment (\textit{standard}) as our reference for comparison. The measured monopole power spectrum is shown in the bottom panel. This exercise illustrates that the four known maps are not sufficient to eliminate the systematic trend in the mean density against the Gaia stellar map. We also show that incorporating the SDSS stellar map (\textit{NN known+sdss}) and even all of the SDSS maps (\textit{NN all sdss}) are not adequate to obtain satisfactory cleaning. On the other hand, we note that \textit{NN all sdss} performs as well as \textit{Lin Gaia}, implying that the nonlinear nature of NN can mitigate the effect of the missing input to some extent. Interestingly, adding the SDSS stellar map to the four known maps impacts the result adversely, especially on the low density end of the Gaia DR2 stars. This result might mean that the SDSS stellar density map is not a proper proxy of the stellar contamination effect in the regions with a lower stellar density. This test indicates that the Gaia DR2 stellar map is proved pivotal to perform a robust cleaning of data, that is consistent with the statistical tests of the mocks. For comparison, we apply the linear model with MSE and the additional Gaia map (\textit{LIN known+gaia}), and obtain a total chi-squared value of $\chi^{2}=196.9$ which is still significant (see Table \ref{tab:chi2methods}). This test demonstrates that capabilities of linear treatment is limited even after including the Gaia map, and the nonlinear treatment is crucial, not  only when we know the proper set of the input templates a priori, but also when we do not know. 

\subsubsection{Redshift Slicing and Pixel Resolution}\label{subsec:slicing}

\begin{figure}
    \centering
    \includegraphics[width=0.45\textwidth]{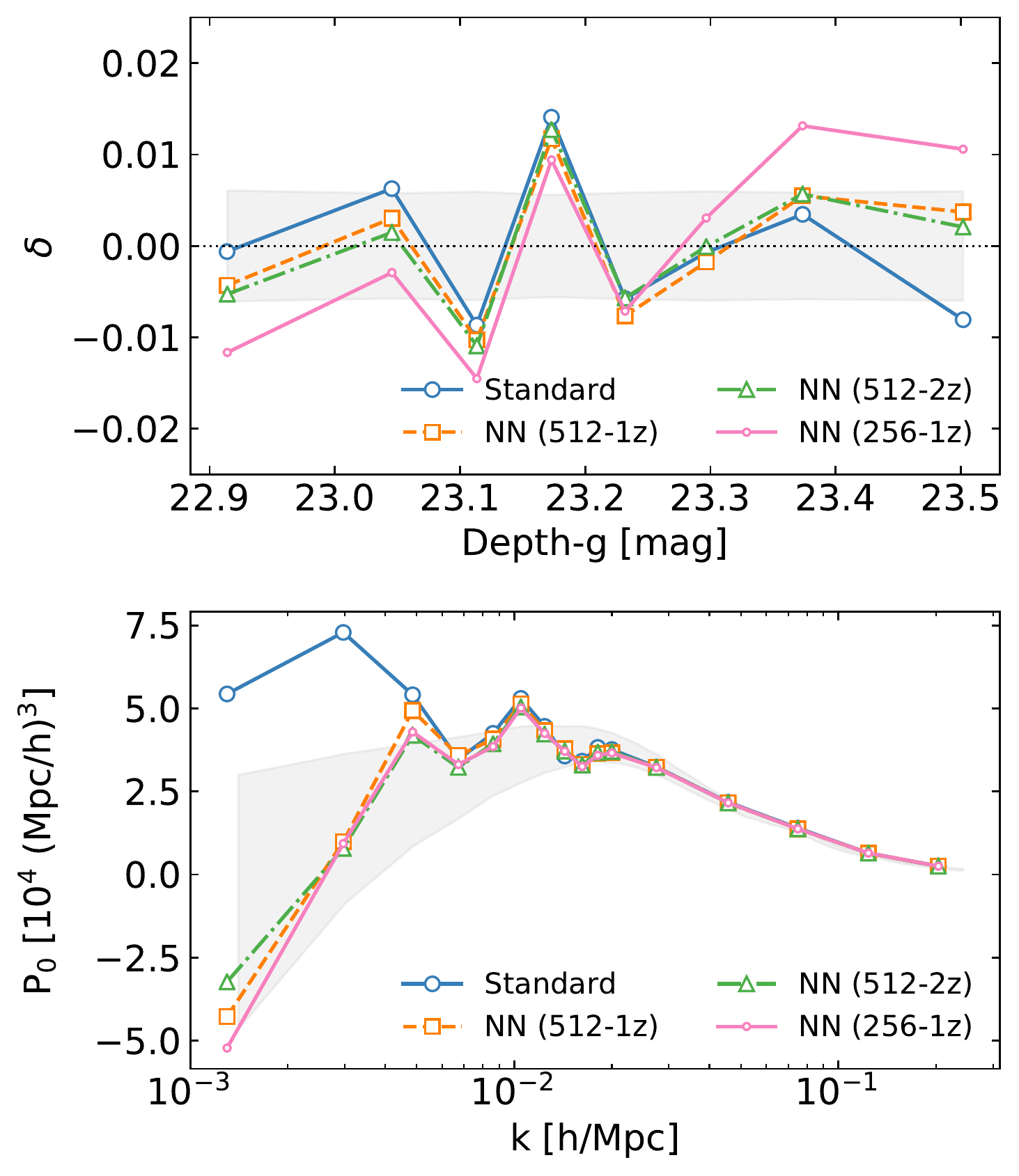}
    \caption[Mean density and power spectrum of eBOSS quasars for different resolutions.]{\textit{Top}: Mean density contrast of the NGC main sample against depth in the g-band after training the neural network methods with two redshift slices with \nside=512 (\textit{NN-512-2z}), one redshift slice with \nside=256 (\textit{NN-256-1z}), and one redshift slice with \nside=512 (\textit{NN-512-1z}). \textit{Bottom}: Monopole power spectrum after the same techniques. The shades represent 1$\sigma$ statistical uncertainty constructed from the null EZmock realizations.}
    \label{fig:nbarsplit}
\end{figure}

The effects of imaging systematics on target density might vary along the line of sight, and alter the slope of the redshift distribution of quasars as well as its overall magnitude. This effect can be ideally investigated by slicing the sample into smaller redshift bins to construct a 3D selection function of imaging systematics. However, further splitting the sample increases the sparsity and makes it difficult to separate the effects of imaging systematics and noise. Thus, we begin with splitting the main sample in the NGC with $\textsc{nside}=512$ into $0.8<z<1.5$ and $1.5<z<2.2$ and then we perform regression on each slice separately. We use PNLL as the cost function with the four known maps and the Gaia stellar map as input. Fig. \ref{fig:nbarsplit} compares the mean density contrast as a function of depth in g band (\textit{top}) and the measured monopole power spectrum (\textit{bottom}) after this treatment (\textit{NN 512-2z}) with that of the standard method and the neural network, trained on $0.8<z<2.2$ with \nside=512 (\textit{NN 512-1z}). This plot demonstrates that there is no evidence for redshift-dependent systematic effects due to imaging.

We also train the neural network with coarser imaging templates in \nside=256 on $0.8<z<2.2$ (\textit{NN 256-1z}). The mean density contrast and measured power spectrum after this treatment are shown in Fig. \ref{fig:nbarsplit} in the top and bottom panels, respectively. The mean density contrast histogram is computed against the depth template in \nside=512 and indicates noticeable residual variations from zero in the regions at low and high \textit{depth-g}. This test demonstrates that the systematic weights obtained in the lower resolution cannot properly reduce the trends in the higher resolution. As a caveat, if the top panel was drawn against the depth template in \nside=256, we would have gotten a reasonable performance of NN-256. The bottom figure shows the resulting power spectrum accounting for the effects in all 17 imaging maps. We find a reasonable stability against the varied pixel resolutions and choose NN-512 as our default.

The total $\chi^{2}$ of the mean density residuals for various systematic treatment methods are summarized in Table \ref{tab:chi2methods}. The $\chi^{2}$ value is $1344.9$ before correcting for systematic effects. After using the linear cleaning techniques, the $\chi^{2}$ value drops below $220$. The non-linear approach lowers the error below 200, and changing the cost function from MSE to PNLL improves the performance even further by returning a value around $170$, which is less than what is observed in the null EZmock realizations. As a comparison, purely cosmological signal without systematics, estimated from the null EZmocks, returns the $\chi^{2}$ value of 178.

\begin{table}
    \centering
    \caption[Total chi-squared value of eBOSS quasars for various methods.]{Total $\chi^{2}$ of the mean density residuals of the main sample in the NGC after various mitigation configurations. The chi-squared value before accounting for imaging systematics is $1344.9$, which is much larger than the 95-th quantile observed in the null EZmocks, i.e., $178$.}
    \resizebox{0.46\textwidth}{!}{
    \begin{tabular}{c|lcccc}
    \hline 
    & && \textit{templates} && \\
    \cline{3-6}
    \nside-\textit{Split}&& \textit{known} &	\textit{known+SDSS} &	\textit{all SDSS} &	\textit{known+Gaia} \\
    \hline 
    \multirow{6}{*}{512-$1z$}
                                       & standard &218.1 &- &-&-\\
                                       & linear-mse &213.5&-&-&196.9\\
                                       & linear-pnll &210.2&-&-&-\\
                                       & nn-mse &194.6&-&-&- \\
                                       & nn-pnll-lr &\textbf{168.99}&-&-&- \\
                                       & nn-pnll &\textbf{163.97}&184.6&\textbf{153.9}& \textbf{151.7}\\
    \hline
    512-$2z$ &	nn-pnll &	- & - & - & \textbf{165.5}\\
    \hline
    256-$1z$ & nn-pnll & - & - & - & 217.6\\
    \end{tabular}
    }
    \label{tab:chi2methods}
\end{table}

\subsection{Significance of Residual Fluctuations}\label{subsec:sigreserror}
For testing the significance of residual systematics, and as our default NN approach, we focus on the neural network method trained with PNLL and cyclic learning rate on the DR16 sample covering $0.8<z<2.2$ with the five imaging maps as input features. From the previous analyses, see Table \ref{tab:chi2methods}, we identify this configuration as the optimal approach. 

\begin{figure*}
  \centering
  \includegraphics[width=0.99\textwidth]{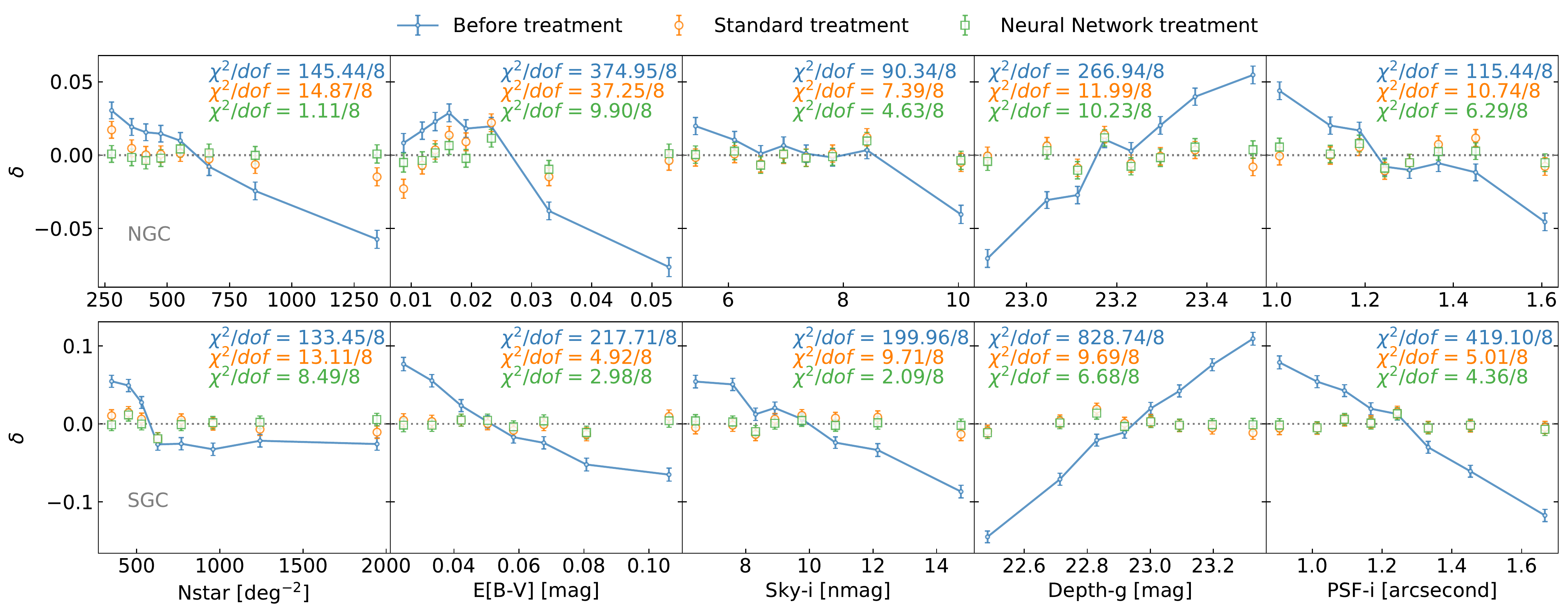}
  \caption[Density of eBOSS quasars as a function of imaging properties.]{Density contrast of the main sample as a function of the primary imaging properties in the NGC (top) and SGC (bottom) before and after accounting for systematic effects using the standard method or neural network. The error-bars are estimated from the null EZmocks and used to calculate the $\chi^{2}$ of the mean density residuals against each imaging property.}
  \label{fig:data_nbar}
\end{figure*}

Fig. \ref{fig:data_nbar} illustrates the observed density contrast of the DR16 quasars as a function of imaging properties for the NGC (top) and SGC (bottom). Respectively from left to right, the imaging quantities are the Gaia DR2 stellar density, Galactic extinction, sky brightness in i band, depth in g band, and seeing in i band. Each panel is annotated with the residual squared errors that are calculated against a zero model as the ground truth\footnote{We assume that the density contrast must be zero when averaged over many pixels in the absence of imaging systematics.}. The error bars are obtained from the EZmock catalogues. We observe the biggest variation is against the extinction for about $8\%$ with $\chi^{2}/{\rm dof}=374.95/8$ in the NGC and $15\%$ against depth-g with $\chi^{2}/{\rm dof}=828.74/8$ in the SGC. Interestingly, the neural network treatment is capable of modeling and removing the non-linear effects in the sample. On the other hand, the standard treatment leaves a significant chi-squared value against the extinction with $\chi^{2}/{\rm dof}=37.25/8$ in the NGC.

\begin{table}
    \centering
    \caption[Total $\chi^{2}$ for eBOSS quasars for North and South Galactic caps.]{Total $\chi^{2}$ of the mean density residuals for the main and high-z samples, before and after mitigation, and the 95-th percentile of null EZmocks. The null EZmock covariance matrix is used to calculate these statistics.}
    \begin{tabular}{c|ccccc}
    & & Noweight & Standard & NN & 95-th \%\\
    \hline 
    \multirow{2}{*}{\textbf{NGC}} & $0.8<z<2.2$ & 1344.9 & 218.1 & 151.7 & 178.0\\
                                  & $2.2<z<3.5$ & 1752.0 & 121.6 & 104.2 & -- \\
        \hline
        \multirow{2}{*}{\textbf{SGC}}& $0.8<z<2.2$ & 1943.0 & 132.5 & 116.3 & 179.2\\
                            & $2.2<z<3.5$ & 2553.7 & 146.0 & 130.1 & --
    \end{tabular}    
    \label{tab:chi2_nbar}
\end{table}

\begin{figure*}
    \centering
    \includegraphics[width=0.45\textwidth]{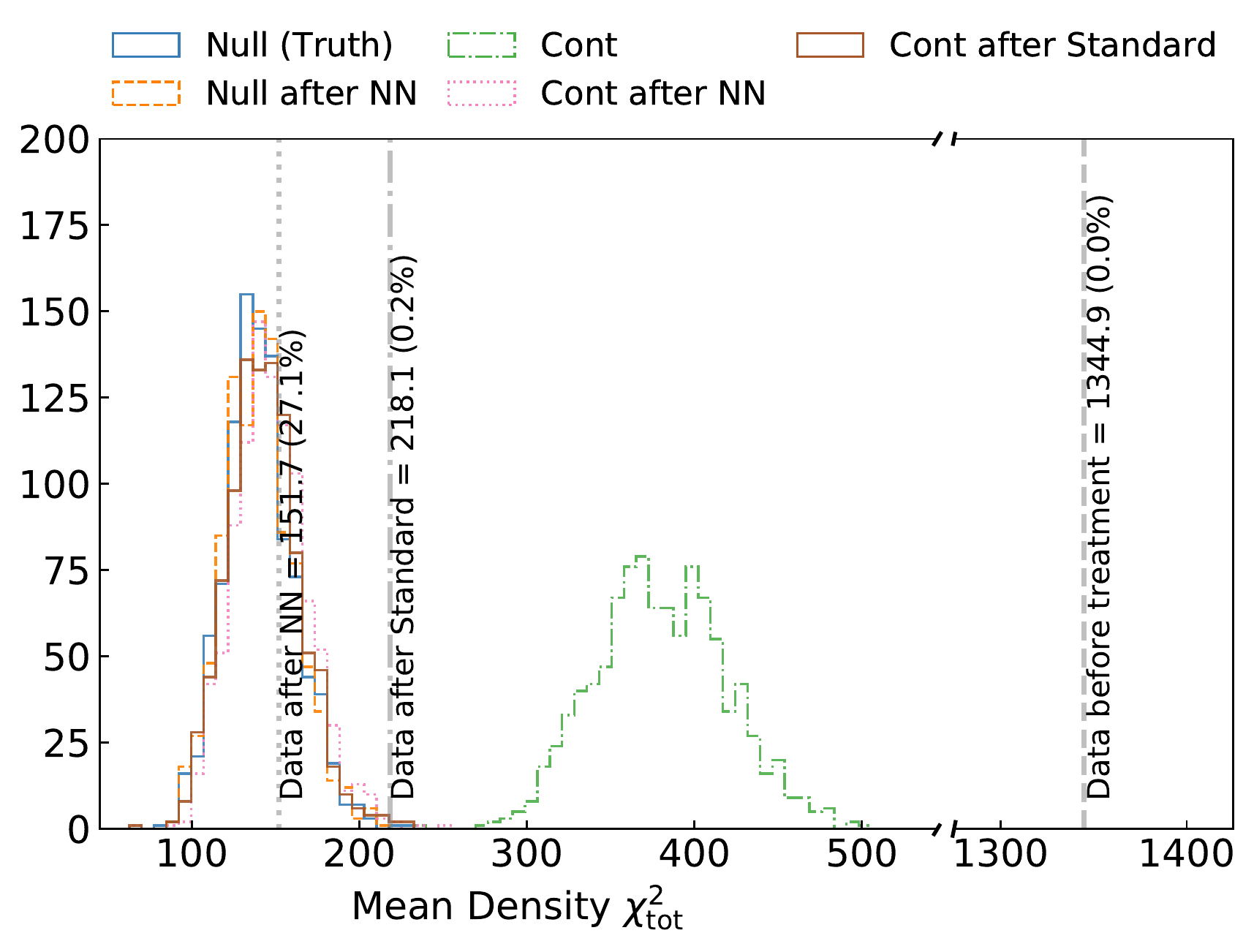}
    \includegraphics[width=0.465\textwidth]{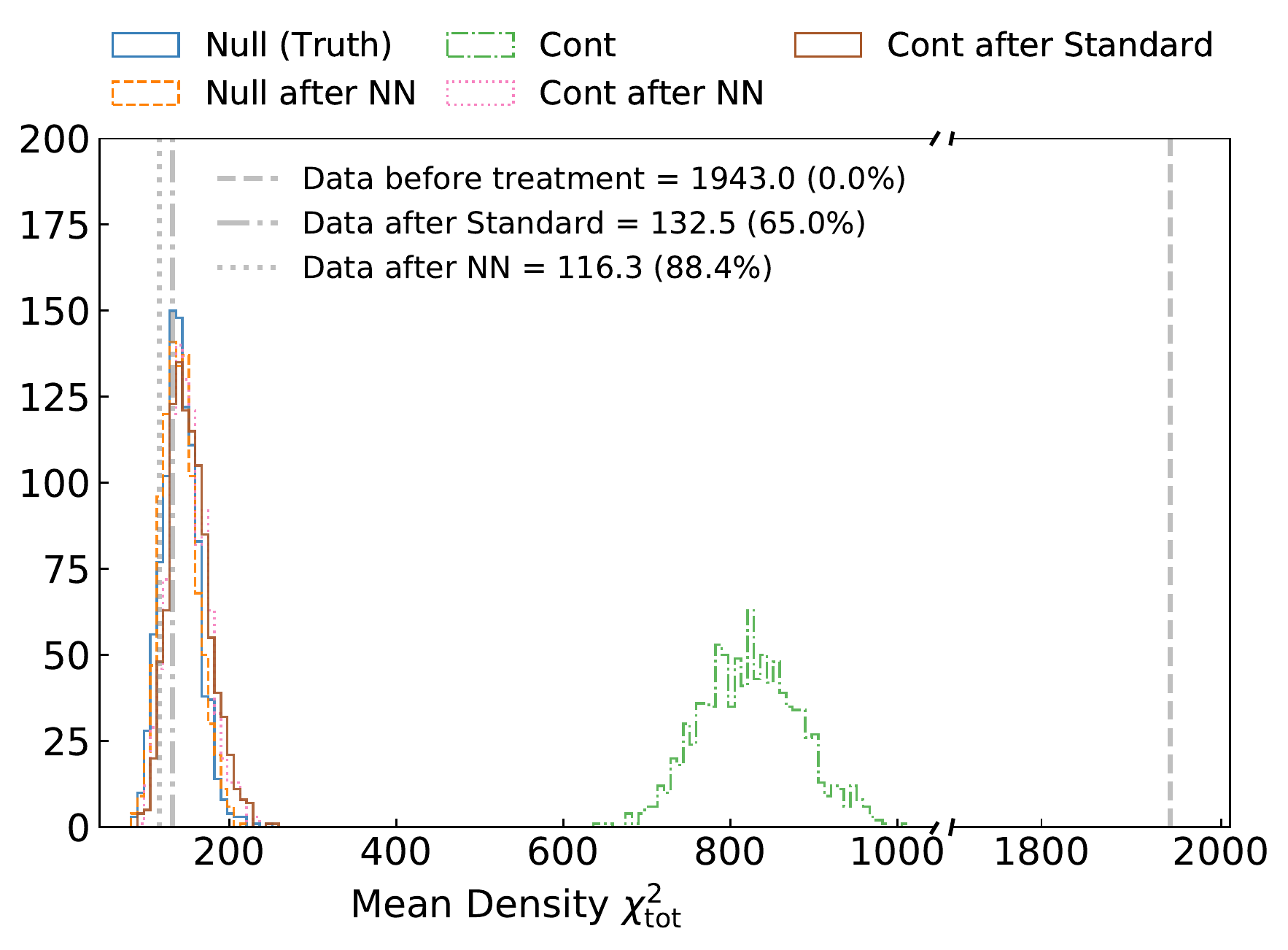}
    \caption[Total chi-squared value of eBOSS quasars from mean density contrast.]{Total $\chi^{2}$ of the mean density residuals for the EZMocks with systematics (Cont) and without systematics (Null) for the NGC (left) and SGC (right). The same statistics observed in the eBOSS DR16 quasar sample (before and after treatment) are shown via vertical lines with the associated p-values, which are derived by comparing with the mocks without systematics, Null (Truth). There is substantial remaining systematics in the NGC with the standard linear treatment. Also, the contaminated simulations do not reflect the same level of systematic effects as the DR16 sample.}
    \label{fig:chi2_nbar}
\end{figure*}

We compute the total $\chi^{2}$ of the mean density residuals against all of the 17 imaging maps to determine the significance of spurious fluctuations in the observed mean density of quasars. Fig. \ref{fig:chi2_nbar} shows the distributions of $\chi^{2}_{\rm tot}$, which are constructed from the null (Null) and contaminated EZmocks (Cont), before and after applying imaging systematics mitigation for the NGC (left) and SGC (right). The values observed in the DR16 sample before and after cleaning are represented with vertical lines. We use the distribution of the null mocks (Null Truth) to compute $p{\rm -value}$. In the NGC, the standard treatment yields $\chi^{2}=218.1$ with $p{\rm -value}=0.2\%$, while the neural network treatment cleans the sample substantially and returns a smaller $\chi{2}$ and higher $p{\rm -value}$, respectively, $151.7$ and $27.1\%$. In the SGC, we observe that the standard method returns $132.5$ with $p{\rm -value}=65.0\%$. This residual is somewhat expected since the trends against imaging maps in the SGC are mostly linear, and thus a linear model is sufficient for cleaning (see Fig. \ref{fig:data_nbar}). In the SGC, both methods return statistics that are in agreement with the $\chi^{2}$ distribution of the null mocks. No remaining systematic error is observed within the statistical uncertainty of the mocks. The total $\chi^{2}$ of the mean density residuals for the main and high-z samples are reported in Table \ref{tab:chi2_nbar}. The 95-th percentile for the main sample is estimated from the mocks and reported in the last column. Note that for the NGC region, the standard approach yields a $\chi^{2}$ value that is larger than the 95-th quantile of the mocks.

\begin{figure*}
    \centering
    \includegraphics[width=0.455\textwidth]{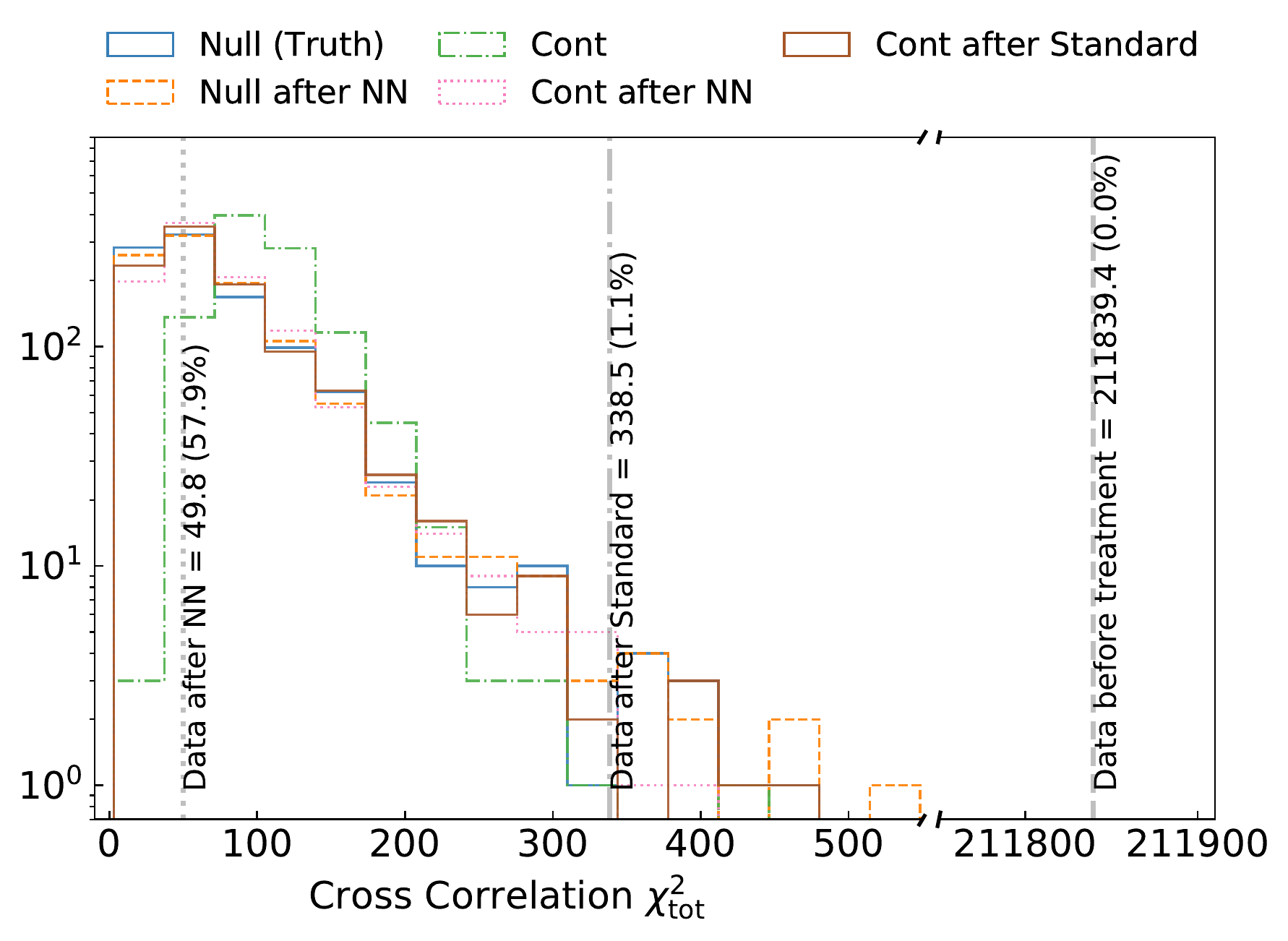}
    \includegraphics[width=0.45\textwidth]{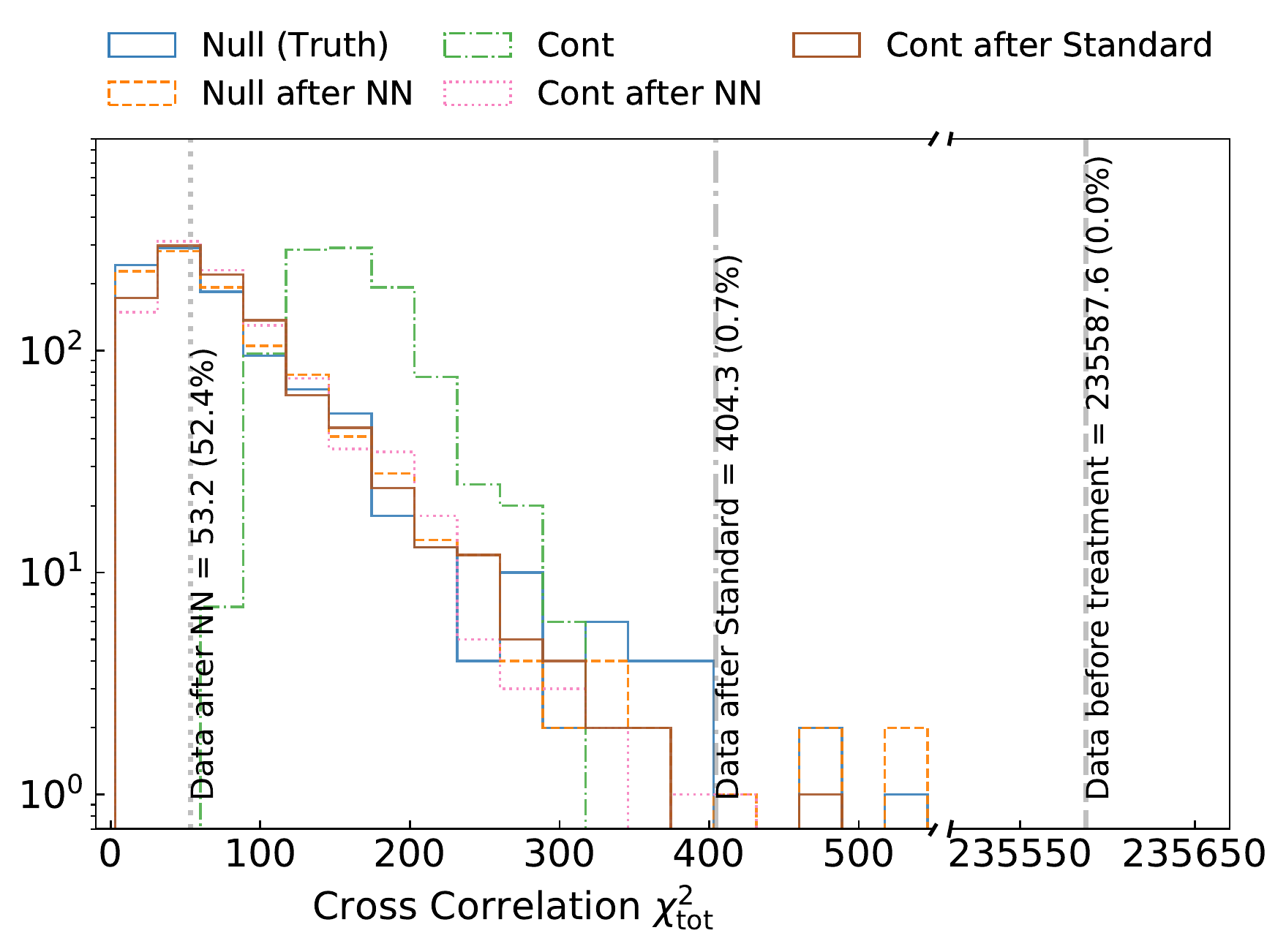}
    \caption[Total chi-squared value of eBOSS quasars from cross power spectra.]{Similar to Fig. \ref{fig:chi2_nbar} for the angular cross power spectrum between the projected quasar density and imaging maps. Compared with the simulations without systematics (Null), there is a substantial remaining systematics both in the NGC and SGC with the standard linear treatment, while the 1D diagnostic (Fig. \ref{fig:chi2_nbar}) shows no obvious issues with the SGC sample cleaned with the linear approach.}
    \label{fig:chi2_cell}
\end{figure*}

Similarly, we cross-correlate the quasar density map with all of the 17 imaging templates. The cross-correlations are then binned to decrease statistical fluctuations, and normalized by the auto power spectrum of the imaging maps. The first four bins are then used to compute the residual squared error against zero. Fig. \ref{fig:chi2_cell} demonstrates the distribution of $\chi^{2}_{\rm tot}$ constructed from the null and contaminated mocks before and after applying imaging systematics mitigation for the NGC (left) and SGC (right). The values observed in the DR16 sample are represented with vertical lines. In the NGC, the standard treatment returns $\chi^{2}=338.5$ with $p{\rm -value}=1.1\%$, while the neural network treatment provides a cleaner sample with $\chi{2}=49.8$ and $p{\rm -value}=57.9$, respectively. In the SGC, we observe that the standard method is incapable of removing the systematics by returning $\chi^{2}=404.3$ with $p{\rm -value}=0.7\%$. On the other hand, the neural network approach enables rigorous cleaning with $\chi^{2}=53.2$ and $p{\rm -value}=52.4\%$. This test motivates further investigations of linear systematic treatment methods in future galaxy surveys since the 1D diagnostic based on the mean density contrast is not sufficiently sensitive to unveil these issues with the standard treatment in the SGC (see Fig. \ref{fig:chi2_nbar}). Interestingly, these histograms show that the magnitude of simulated systematic effects for the contaminated mocks are stronger in the SGC (cf. the left and right panels of Fig. \ref{fig:chi2_nbar} and \ref{fig:chi2_cell}). Similarly, the DR16 sample shows a stronger spurious fluctuation around $15\%$ against depth in the SGC region, compared with $8\%$ in the NGC.

In summary, we find that the nonlinear aspect of our cleaning approach is the primary reason for efficiently reducing spurious fluctuations and systematic error. The neural network-based approach can model the non-linear feedback of observed quasar density to imaging templates, which results in a cleaner sample with a significantly lower $\chi^{2}$ value. Although the standard treatment passes the null test based on mean density contrasts, however, the test based on cross power shows that the catalog with the standard systematic weights is not properly cleaned. We find no improvement in the mean density residual after including all SDSS maps for training; however, our analysis shows that the Gaia stellar density is required to satisfy the null tests for residual systematic errors. We also find that computing the mean quasar density per pixel does not change our conclusion based on the mean quasar density per imaging bin, although the former quantity is subject to more fluctuations. Finally, we do not observe a significant change by splitting the main sample into redshift subsamples or using coarser imaging templates. In the following, our neural network-based treatment is applied on the entire $0.8<z<2.2$ and uses the Poisson cost function (nn-pnll), cyclic learning rate, and five imaging templates in \textsc{nside}$=512$ (Known+Gaia) as input features, see, Tab. \ref{tab:chi2methods}.

\subsection{Power Spectrum}
\subsubsection{Measurement}
\begin{figure*}
    \centering
    \includegraphics[width=0.45\textwidth]{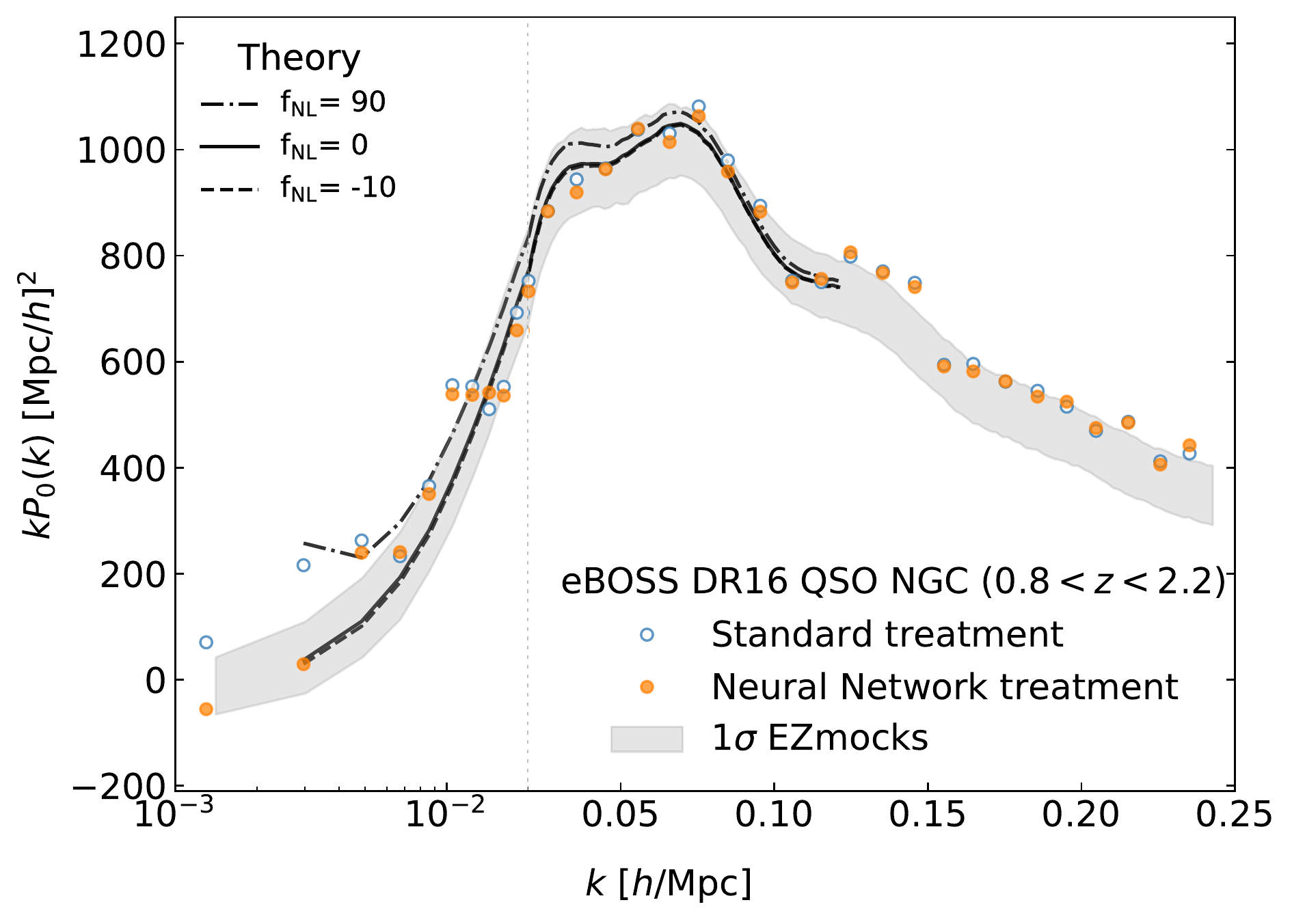}
    \includegraphics[width=0.45\textwidth]{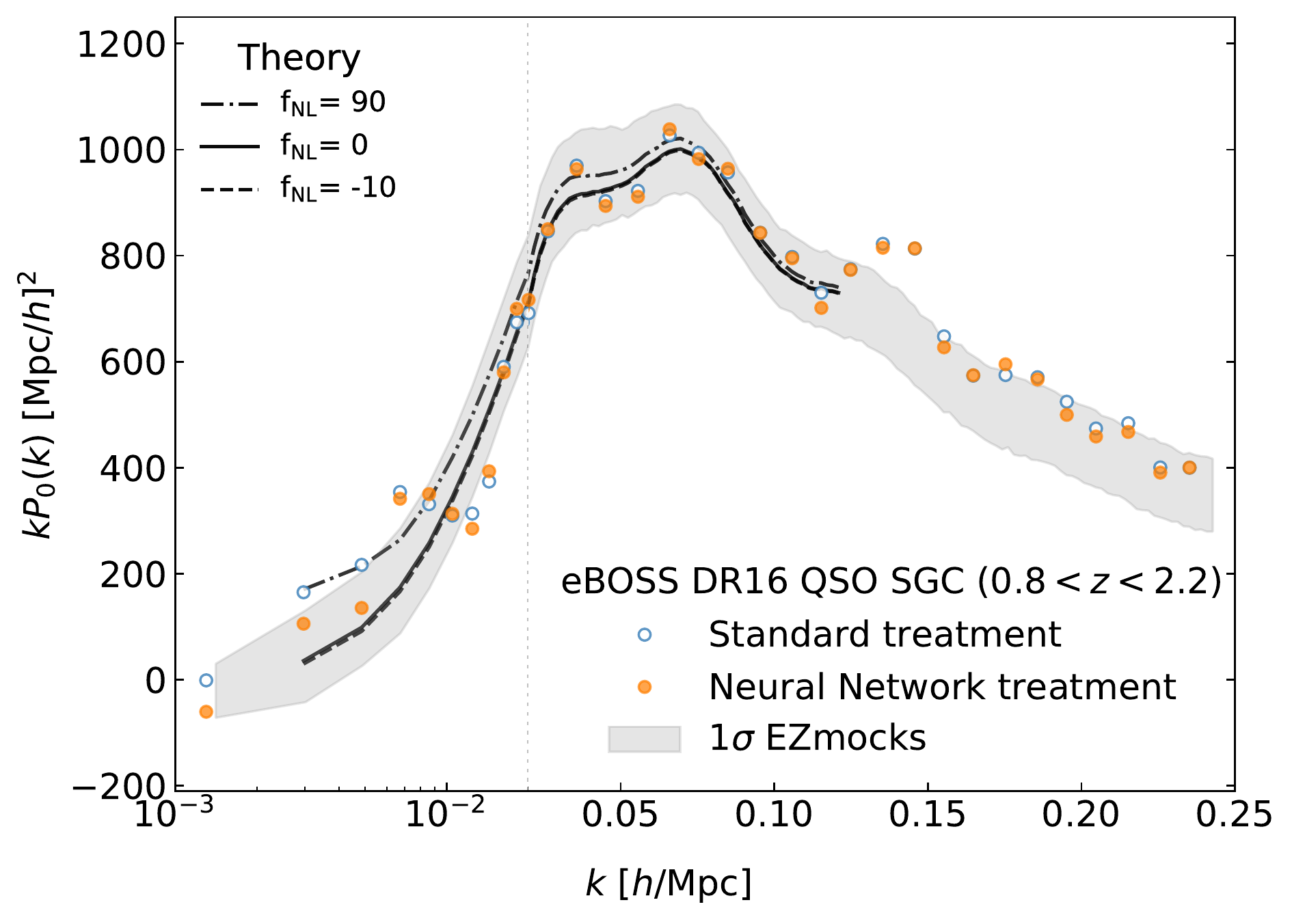}
    \caption[Power spectrum of eBOSS quasars in the NGC and SGC.]{Monopole of the main sample in the NGC (left) and SGC (right) after treatment with the standard method and neural network. Various \fnl~models are plotted to show the sensitivity of the signal on large scales. The shades represent 1$\sigma$ statistical uncertainty estimated from the EZmocks. The x-axes are logarithmic for $k < 0.02~h{\rm Mpc}^{-1}$ and linear otherwise.}
    \label{fig:p0data}
\end{figure*}

We now proceed to measure the power spectrum of the DR16 sample and EZmock simulations for each galactic cap separately since each cap is subject to different targeting properties. Fig. \ref{fig:p0data} shows the measured monopole power spectrum $P_{0}$ of the main sample in the NGC (left) and SGC (right). We use the square-root of the diagonal terms of the covariance matrices, constructed from the null EZmocks, as the errorbars on $P_{0}$. The open and filled circles represent the measured spectrum after cleaning the sample with the standard and neural network treatments, respectively. In both regions, the nonlinear cleaning approach returns a lower power at small $k$. On the other hand, the effect on the small-scale clustering is very small. We also show various models with \fnl=$-10$, $0$, or $90$ to illustrate the sensitivity of the signal on the low-$k$ measurements.

\begin{figure}
    \centering
    \includegraphics[width=0.45\textwidth]{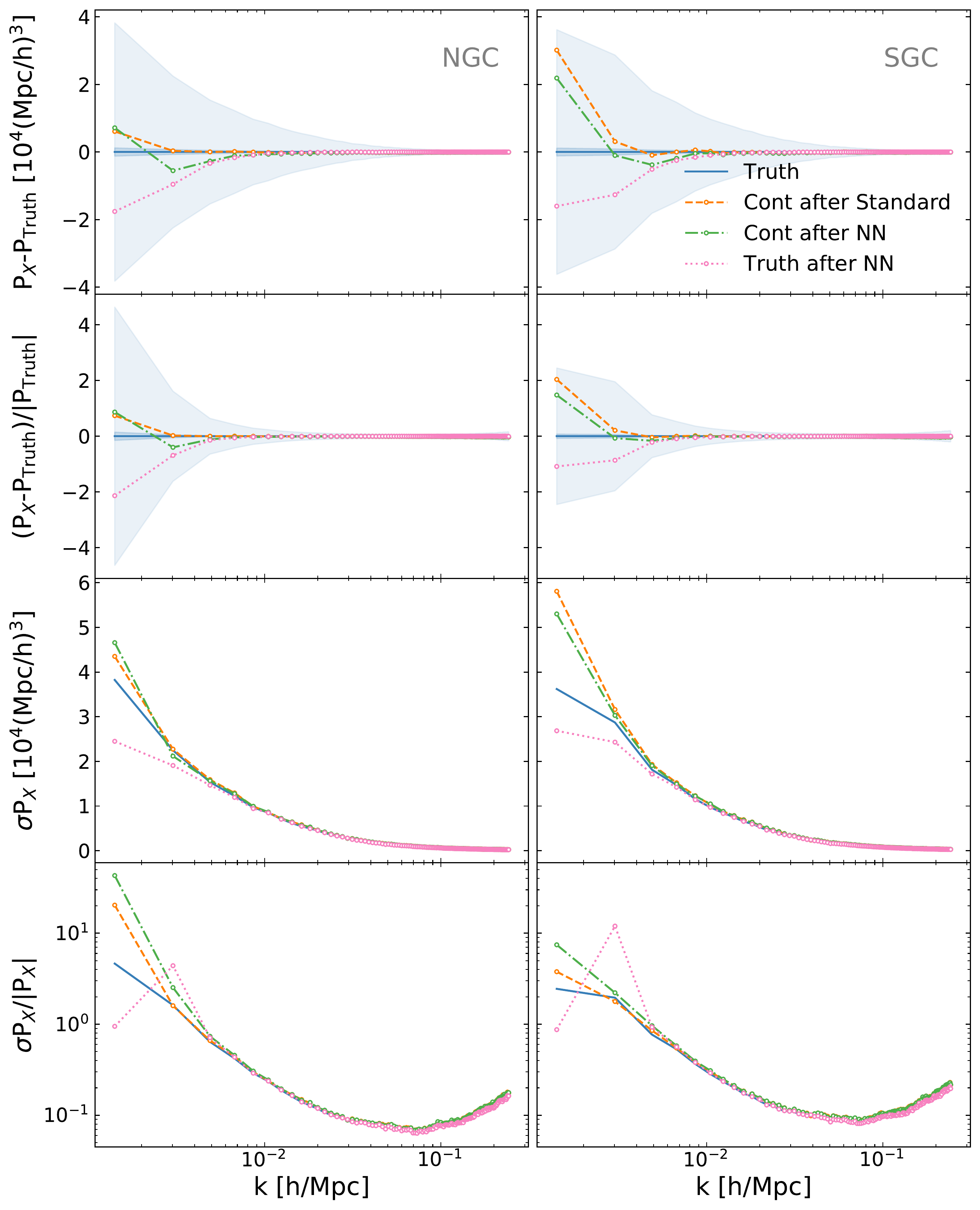}
    \caption[Power spectrum of eBOSS mocks.]{Measured power spectrum of the EZmock realizations before and after systematic treatment for the NGC (left) and SGC (right) regions. From top to bottom, we show the difference between the mean mitigated spectrum and the mean truth spectrum, the relative difference, the dispersion in the mock spectra, and the relative dispersion.}
    \label{fig:p0mocks}
\end{figure}

We apply the systematics treatment methods on both the null and contaminated EZmock catalogues to characterize the impact of the mitigation procedure on the measured clustering statistics. The measured power spectrum of the null mocks without any systematic treatment is considered as the ground truth clustering. Fig. \ref{fig:p0mocks} shows the mean and the standard deviation of the measured spectra from the EZmock realizations in the NGC (left) and SGC (right) regions. The top row illustrates the difference between the mean $P_{0}$ of the EZmocks after mitigation, including the null (\textit{Truth after NN}) and contaminated catalogues using the neural network  (\textit{Cont after NN} and the standard approach \textit{Cont after Standard}), and the truth clustering (\textit{Truth}). The light and dark shades show the standard deviation of the null EZmock spectra and the $1\sigma$ uncertainties on the mean of the mock spectra, respectively. In the second row, we show the relative difference in the mean power spectrum. In the third row, we present the standard deviation of the mock spectra. Finally, we show the relative dispersion in the bottom row. The measured spectra for the contaminated mocks before treatment is an order of magnitude larger than the truth clustering and thus is not visualized for clarity. We note that the magnitude of the excess power observed in the mocks is one order of magnitude smaller than what is observed in the real sample (cf. Fig. \ref{fig:p0data}), primarily because a linear model was used to generate systematics. This implies that the actual systematics of the real sample is substantially more severe and complex than in these mocks.

Due to allowing the correction to account for more freedom, the neural-network treatment removes more of the modes, known as the over-fitting problem, when it is applied to the mocks. On the other hand, the standard treatment indicates less of this over-fitting issue. This is expected as the same linear model is used to produce the systematic effects in the mock realizations. The standard deviation of the mock spectra shows that the imaging treatments do not increase the fractional variance of the measured power spectrum down to $k=0.003~h/{\rm Mpc}$. Interestingly, we observe that the dispersion of the null mocks decreases after applying the NN treatment, which is due to the over-correction of the power itself \footnote{The error on the power spectrum is expected to be proportional to the power itself under a Gaussian limit.}.

\subsubsection{Mitigation Bias}
Using the mocks, we attempt to estimate any residual or over-correction that might have been introduced in the measured eBOSS QSO power spectrum in the process of the imaging systematics treatment. This assessment is crucial for obtaining an unbiased clustering measurement, which will lead to an accurate inference of cosmological parameters.

\begin{figure}
    \centering
    \includegraphics[width=0.45\textwidth]{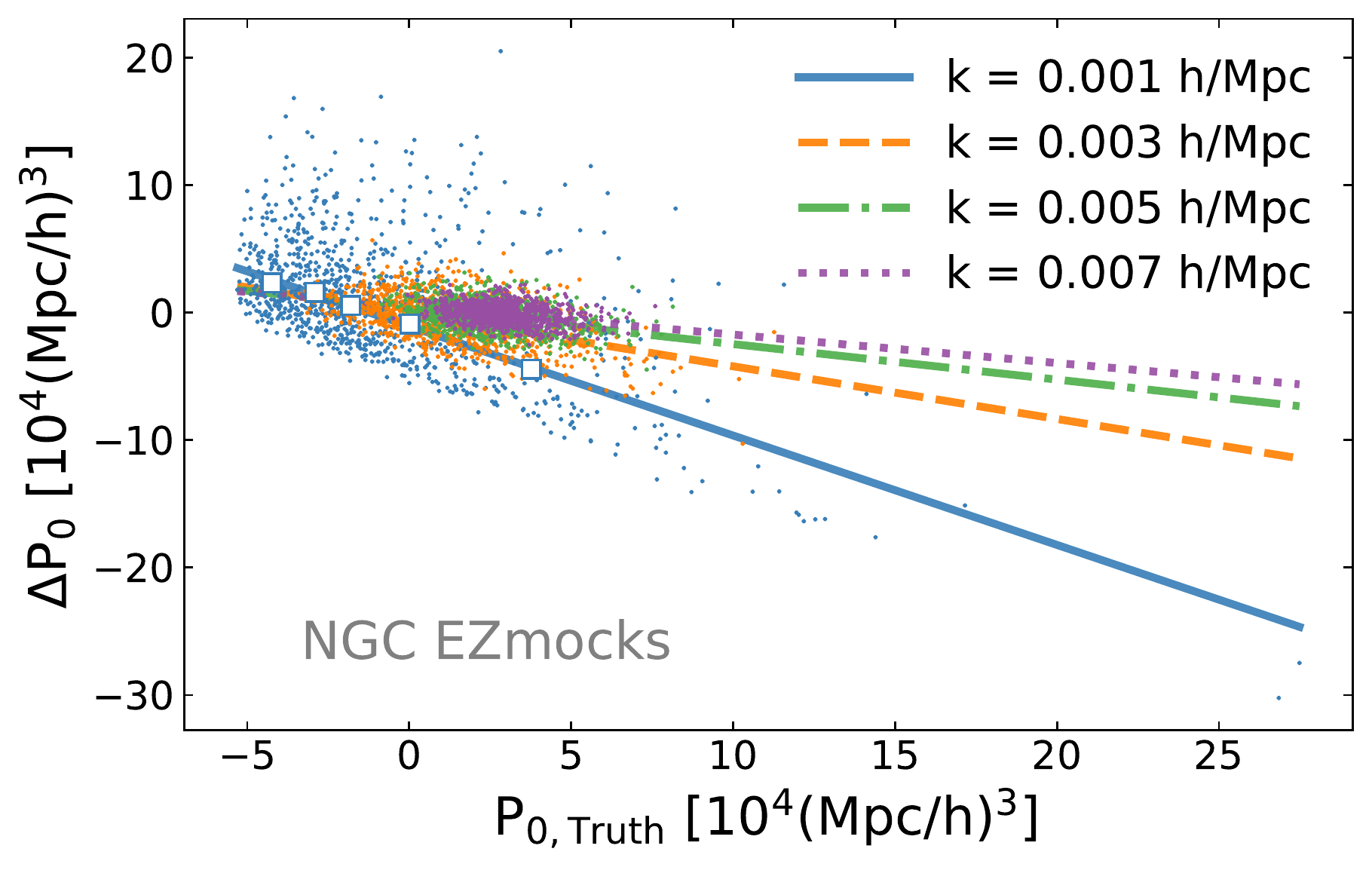}
    \caption[Residual bias between mock power spectra after mitigation.]{Difference between the measured spectra of the contaminated EZmock catalogues after cleaning and the spectra of the null catalogues as a function of the latter. The medians are used to obtain the best linear fit in each k bin, and are shown only for $k=0.001~h/{\rm Mpc}$ with open squares.}
    \label{fig:dpvsp}
\end{figure}

We compare the spectra of the contaminated mocks after mitigation to that of the null mocks before treatment as the true power. Fig. \ref{fig:p0mocks} shows that the NN-based mitigation tends to introduce overcorrection for $k< 0.003$, especially if it is applied on the density field with no systematics (i.e., Truth with NN.~\footnote{The standard method performs better by construction, as it assumes we know exactly the source of systematics.}). With this caveat, we focus on the overcorrection observed in the contaminated mocks and inspect its nature. The differences between the measured power of the 1000 contaminated mocks after cleaning and the true power for all mocks are shown in Fig. \ref{fig:dpvsp} as a function of the true power for the first few $k$ bins. We find that the mitigation bias (or overcorrection) is almost linearly proportional to the true power. While this exercise shows that if the data has no systematics or is subject to simple, linear systematics, the neural network based method we developed will potentially introduce a small degree of overcorrection and we can attempt to correct for such mitigation bias. However, from Fig. \ref{fig:chi2_nbar} and \ref{fig:chi2_cell}, it is apparent that the DR16 sample is subject to much more severe and nonlinear systematic effects compared to the mocks and the standard linear method is not sufficiently effective. Therefore, we believe that the overcorrection is less likely a problem for the real eBOSS data and an attempt to mitigate it might further bias the data. We present the discussion of our mitigation bias on primordial non-Gaussianity constraints in a companion paper \citep{mueller2020fnl}.

\section{Conclusion}\label{sec:conclusion}
We have performed a thorough study of imaging systematic effects and various template-based mitigation techniques in the final sample of quasars \citep{lyke2020dr16qso, ross2020lss} from the eBOSS DR16 \citep{Ahumada2020ApJS}. We present a nonlinear cleaning approach, based on artificial neural networks, and compare the treatment effectiveness with the standard method, based on linear regression. The methods are applied to model the observed density of quasars given a set of templates for imaging properties, related to SDSS properties and Galactic foregrounds, which include stellar density, Galactic extinction, neutral hydrogen column density, depth, seeing, sky brightness, airmass, and run. As summarized in Tab. \ref{tab:chi2methods},
\begin{enumerate}[leftmargin=1\parindent]
    \item We find that the neural network-based approach outperforms standard linear regression by allowing more freedom for correcting nonlinear and complex variations in the quasar density caused by imaging properties, see Fig. \ref{fig:nbarmethods} and \ref{fig:data_nbar}. The approach is also further improved by using the Poisson statistics to account for the sparsity of the DR16 sample.
    
    \item Stellar density is one of the most important sources of spurious fluctuations, and a new template constructed using the Gaia DR2 \citep{gaia2018} yields the best agreement to the observed chi-squared values in the simulations, see Fig. \ref{fig:nbarnstar} and Tab. \ref{tab:chi2methods}. We also show that linear treatment, with the Gaia map included, is still not able to properly remove systematics.
    
    \item We find no evidence for redshift-dependent imaging systematics and no substantial difference after changing the pixel resolution of imaging templates, see Fig. \ref{fig:nbarsplit}. Therefore we choose NN trained with PNLL, cyclic learning, and imaging templates in $\nside=512$ as our default approach.
    
\end{enumerate}

We utilize the EZmocks, both in the presence and absence of imaging systematics, to construct covariance matrices, quantify residual systematic error, and assess the quality of the DR16 sample for cosmological studies. We find
\begin{enumerate}[resume, leftmargin=1\parindent]

    \item The mean density null test shows some remaining systematic error in the catalogue with the standard weights in the NGC region, specifically $\chi^{2}=218.1$ with $p{\rm -value}=0.2\%$. Although this test does not reveal any issues with the standard catalogue in the SGC region, $\chi^{2}=132.5$ with $p{\rm -value}=65.0\%$ (see Tab. \ref{tab:chi2_nbar}), our second null test based on cross-power spectra unveils a significant systematic error in the SGC, $\chi^{2}=404.3$ with $p{\rm -value}=0.7\%$.
    
    \item This work motivates further investigations of linear systematic treatment methods in future galaxy surveys since the 1D diagnostic based on the mean density contrast does not indicate any issues with the standard treatment in the SGC (see Fig. \ref{fig:chi2_nbar}).
    
    \item The catalogue with the neural network-based systematic weights passes both null tests by providing substantially lower $\chi^{2}$ values, see Fig. \ref{fig:chi2_nbar} and \ref{fig:chi2_cell}.
    
\end{enumerate}

Collectively, these tests demonstrate that the DR16 quasar catalogue with the standard systematic weights suffer from residual imaging systematics in both Galactic caps, and should not be used for measuring quasar clustering on large scales, i.e., $k < 0.01~h/{\rm Mpc}$, as shown in Fig. \ref{fig:p0data}. Nevertheless, it is expected that the impact of imaging systematics to be insignificant on the BAO measurements \citep[e.g., analyses presented in ][]{hou2020qso, neveux2020qso}, and a thorough investigation is conducted in a companion paper \citep{merz2020bao}.

We then apply our methods on the EZmocks to quantify the impact of systematics treatment on quasar clustering measurements, see Fig. \ref{fig:p0mocks}. The neural-network treatment removes some of the cosmological power due to allowing for more freedom in removing systematic effects. We find that the impacts of overfitting on the mean of the mock power spectra and its error are marginal for $k > 0.004~h/{\rm Mpc}$. We employ linear regression to model the impact of mitigation on recovering the ground truth clustering, see Fig. \ref{fig:dpvsp}. We emphasize that the utility of the mitigation bias treatment is not clear since the parameters are derived from the mocks without realistic imaging systematics, see Fig. \ref{fig:chi2_nbar} and \ref{fig:chi2_cell}. However, we investigate the effect on primordial non-Gaussianity constraints in a companion paper \citep{mueller2020fnl}.

The end-product from this work is a new value-added quasar catalogue with the enhanced weights to correct for nonlinear imaging systematic effects. The new weights are necessary to make a robust measurement of quasar clustering on large scales ($k < 0.01~h/{\rm Mpc}$). This catalogue is used in a companion paper constraining the local-type primordial non-Gaussianity \citep{mueller2020fnl}.

\section*{Acknowledgements}
We thank the anonymous referee for their insightful comments and suggestions. M.R. is supported by the U.S.~Department of Energy, Office of Science, Office of High Energy Physics under DE-SC0014329; H.-J.S. is supported by the U.S.~Department of Energy, Office of Science, Office of High Energy Physics under DE-SC0014329 and DE-SC0019091. We acknowledge the support and resources from the Ohio Supercomputer Center \citep[OSC;][]{Owens2016}. Specifically, this work utilized more than $359000$ core hours of the Owens cluster. M.R. is grateful for help from Xia Wang, Antonio Marcum, and Yu Feng. G.R. acknowledges support from the National Research Foundation of Korea (NRF) through Grants No. 2017R1E1A1A01077508 and No. 2020R1A2C1005655 funded by the Korean Ministry of Education, Science and Technology (MoEST). We would like to appreciate the open-source software and modules that were invaluable to this research: Pytorch, Nbodykit, HEALPix, Fitsio, Scikit-Learn, NumPy, SciPy, Pandas, IPython, Jupyter, and GitHub.

Funding for the Sloan Digital Sky 
Survey IV has been provided by the 
Alfred P. Sloan Foundation, the U.S. 
Department of Energy Office of 
Science, and the Participating 
Institutions. SDSS-IV acknowledges support and 
resources from the Center for High 
Performance Computing  at the 
University of Utah. The SDSS 
website is www.sdss.org. This work also relied on resources provided to the eBOSS Collaboration by the National Energy Research Scientific Computing Center (NERSC). NERSC is a U.S. Department of Energy Office of Science User Facility operated under Contract No. DE-AC02-05CH11231.

SDSS-IV is managed by the 
Astrophysical Research Consortium 
for the Participating Institutions 
of the SDSS Collaboration including 
the Brazilian Participation Group, 
the Carnegie Institution for Science, 
Carnegie Mellon University, Center for 
Astrophysics | Harvard \& 
Smithsonian, the Chilean Participation 
Group, the French Participation Group, 
Instituto de Astrof\'isica de 
Canarias, The Johns Hopkins 
University, Kavli Institute for the 
Physics and Mathematics of the 
Universe (IPMU) / University of 
Tokyo, the Korean Participation Group, 
Lawrence Berkeley National Laboratory, 
Leibniz Institut f\"ur Astrophysik 
Potsdam (AIP),  Max-Planck-Institut 
f\"ur Astronomie (MPIA Heidelberg), 
Max-Planck-Institut f\"ur 
Astrophysik (MPA Garching), 
Max-Planck-Institut f\"ur 
Extraterrestrische Physik (MPE), 
National Astronomical Observatories of 
China, New Mexico State University, 
New York University, University of 
Notre Dame, Observat\'ario 
Nacional / MCTI, The Ohio State 
University, Pennsylvania State 
University, Shanghai 
Astronomical Observatory, United 
Kingdom Participation Group, 
Universidad Nacional Aut\'onoma 
de M\'exico, University of Arizona, 
University of Colorado Boulder, 
University of Oxford, University of 
Portsmouth, University of Utah, 
University of Virginia, University 
of Washington, University of 
Wisconsin, Vanderbilt University, 
and Yale University.
 
\section*{Data Availability}
\label{sec:dataavail}

The catalogue described in this article will be made public in Github at \url{https://github.com/mehdirezaie/eBOSSDR16QSOE}. The neural network pipeline utilized in this work is publicly available at \url{https://github.com/mehdirezaie/sysnetdev}.

\bibliographystyle{mnras}
\bibliography{refs} 


\bsp	
\label{lastpage}

\end{document}